\documentclass[12pt]{article}
\usepackage{amsbsy,graphicx}

\catcode`\@=11
\@addtoreset{equation}{section}

\catcode`\@=11
\def\section{\@startsection{section}{1}{\z@}{3.5ex plus 1ex minus
   .2ex}{2.3ex plus .2ex}{\large\bf}}
\newcommand{\beq}{\begin{equation}}
\newcommand{\eeq}{\end{equation}}
\newcommand{\bea}{\begin{eqnarray*}}
\newcommand{\eea}{\end{eqnarray*}}
\newcommand{\beaq}{\begin{eqnarray}}
\newcommand{\eeaq}{\end{eqnarray}}
\def\ba{\begin{array}}
\def\ea{\end{array}}
\def\bi{\begin{itemize}}
\def\ei{\end{itemize}}
\def\bn{\begin{enumerate}}
\def\en{\end{enumerate}}
\def\nn{{\nonumber}}

\begin{document}
\begin{flushright} KIAS-P03004\\hep-th/0301075\end{flushright}

\vskip 1cm
\centerline{\Large \bf 
Thermodynamic Bethe Ansatz}
\centerline{\Large \bf 
for boundary sine-Gordon model}
\vskip 1cm

\centerline{\large Taejun Lee and Chaiho Rim} 
\vskip .5cm
\centerline{\it Department of Physics, Chonbuk National University}
\centerline{\it Chonju 561-756, Korea}
\centerline{\it email: tjun@pine.chonbuk.ac.kr, rim@mail.chonbuk.ac.kr}
\vskip 1cm
\vskip 2cm
\centerline{\bf Abstract}
(R-channel) TBA is elaborated 
to find the effective central charge dependence  
on the boundary parameters 
for the massless boundary sine-Gordon model 
with the coupling constant $(8\pi) /\beta^2 = 1+ \lambda $ 
with $\lambda$ a positive integer. 
Numerical analysis of the massless boundary 
TBA demonstrates that 
at an appropriate boundary parameter range 
(cusp point) there exists a singularity crossing phenomena
and this effect 
should be included in TBA 
to have the right behavior of 
the effective central charge.

\section{Introduction} 
\noindent

The low dimensional quantum system 
such as a quantum wire with boundaries 
is not easy to study in terms of mean field approach
due to large quantum fluctuations.
The system is also strongly affected by the 
the existence of boundaries. 
For example, one needs a good knowledge of the  
the low dimensional quantum field theory
to study the quantum Hall edge tunnelling \cite{pls} .

In this work, the massless 
Tomonaga-Luttinger liquid with boundaries 
is studied motivated by SNS junction super-conductor analysis\cite{css,iaff}. 
This system is summarized in terms of the boundary sine-Gordon model(bSG).
\beq
{\cal A}= 
\int d^2x \,\frac1{4\pi}(\partial_{a}\varphi)^2
-\mu^{(1)}_{B} \int_{y=0} \!\! dx \, 2 
\cos(b(\varphi- \varphi^{(1)}_0))
-\mu^{(2)}_{B} \int_{y=R} \!\! dx \,
2 \cos( b (\varphi-\varphi^{(2)}_0))\,.
\label{massless-sG-action}
\eeq
The free energy dependence of the finite system 
on the boundary parameter 
$\chi = b ( \varphi^{(2)}_0 - \varphi^{(1)}_0)$
is one of the demanding questions.
To do this, we first consider the massive sine-Gordon model 
with boundaries 
and put the bulk mass vanish \cite{FSW}.  

The massive bSG is written as 
\beaq
{\cal A}&=&\int d^2x\left[\frac1{4\pi}(\partial_{a}\varphi)^2
-2\mu(\cos(2 b \varphi)-1)\right]
\nn\\
&&- \mu^{(1)}_{B} \int_{y=0}\!\!dx \,
2 \cos( b (\varphi-\varphi^{(1)}_0))
-\mu^{(2)}_{B} \int_{y=R} \!\! dx \, 
2 \cos( b (\varphi-\varphi^{(2)}_0))\,,
\label{massive-sG-action}
\eeaq
The coupling constant $b^2$ is restricted to be less than 1.   
(Note that  $b^2 $ is scaled by $8\pi$ from the conventional choice 
$\beta^2 = 8\pi b^2$ ).

The bulk sine-Gordon model (SG) \cite{bulk-sG}
belongs to the category of two dimensional 
integrable quantum field theories
and allows an exact treatment of the system.
Integrable quantum systems 
have been studied systematically 
after the pioneering work of Zamolodchikov \cite{sasha}.
The integrability of the bSG 
was also demonstrated in \cite{GhoZam}. 

The scale dependence of the system can be studied
by the method of thermodynamic Bethe ansatz (TBA)
\cite{alyosha,yy}.  Suppose a system lies along the y-axis 
with a  finite size $R$ 
and appropriate boundary conditions are imposed at
each end as in Fig~(\ref{fig1}).
The $x$-direction is periodic and its size $L$ is put to $\infty$ 
in the thermodynamic limit. 

\begin{figure}[ht]
\begin{minipage}[t]{7.0cm}
\begin{picture}(50,150) (-30,-10 )
\put(50,0){\line(0,1) {130}}
\put(50,0){\line(1,0) {50}}
\put(50,0.5){\line(1,0) {50}}
\put(100,0){\line(0,1) {130}}
\put(50,130){\line(1,0) {50}}
\put(50,130.5){\line(1,0) {50}}
\put(47, 65){\tiny$\wedge $}
\put(47, 65.5){\tiny$\wedge $}
\put(97,65){\tiny$\wedge $}
\put(97,65.5){\tiny$\wedge $}
\put(30,70){\footnotesize{($\alpha$) }}
\put(105,70){\footnotesize{($\beta$) }}
\put(105,30){L}
\put(75, -10){R}
\end{picture}
\caption{space with two boundaries:
boundary condition ($\alpha$) and 
boundary condition ($\beta$).} 
\label{fig1}
\end{minipage}
\ $\qquad$ \
\begin{minipage}[t]{7.0cm}
\begin{picture}(50,150) (-40, -40)
\put(0,0){\line(1,0) {130}}
\put(0,0.5){\line(1,0) {130}}
\put(0,0){\line(0,1) {50}}
\put(0,50){\line(1,0) {130}}
\put(0,50.5){\line(1,0) {130}}
\put(130,0){\line(0,1) {50}}
\put(55,53){\footnotesize{$|B_\alpha \rangle$ }}
\put(55,-8){\footnotesize{$|B_\beta \rangle$ }}
\put(-3, 25){\tiny$\wedge $}
\put(-3, 25.5){\tiny$\wedge $}
\put(127,25){\tiny$\wedge $}
\put(127,25.5){\tiny$\wedge $}
\put(-10,25){R}
\put(30,-10){L}
\end{picture}
\caption{space with states:
initial states $|B_\alpha \rangle$ 
and final state $| B_\beta \rangle$.}
\label{fig2}
\end{minipage}
\end{figure}

\noindent
The partition function with these boundaries is given as 
\bea
Z_{\alpha\beta} = \mbox{Tr} e^{-L H_{\alpha\beta}} 
\cong  e^{-L E_{\alpha\beta}(R)}
\eea
for large $L$.
$H_{\alpha\beta}$ is the Hamiltonian of the system of size $R$ 
with boundary ($\alpha$) and ($\beta$). 
$E_{\alpha\beta}(R)$ is the ground state energy 
in the thermodynamic limit,
which depends on the size $R$.

The same system can be viewed as the one with initial state 
$|B_\beta \rangle$ and  
final state $|B_\alpha \rangle$.
In this picture, the bulk is periodic in $L$ 
as in Fig.~(\ref{fig2}).
Then the same partition function is evaluated 
using the periodic Hamiltonian $H$ of the system.
\beq
Z_{\alpha\beta}  \equiv e^{-RL f_{\alpha \beta} (R) }
= <B_\alpha | e^{-RH}|B_\beta > 
= \left(  \sum_{\{A\}} 
\frac{<B_a|A> <A|B_b> e^{-R E_A} }{<A|A> }
\right) \,,
\label{massive-free-energy}
\eeq 
where $f_{\alpha \beta} (R) $ is the free energy density per length
and $\{A\}$ is the complete set of the periodic 
Hamiltonian eigenstates.
The finite size effect of the SG (with boundary)
was analyzed in \cite{FSW,LMSS} for diagonal case
using thermodynamic Bethe ansatz (R-channel TBA). 

In section 2, we present the massive (R-channel) 
TBA for the bulk sine-Gordon model 
with boundary sine-Gordon interaction.
In this analysis, we restrict the coupling constant 
$\lambda = 1 /b^2 -1 \equiv n_b +1$ to a positive integer
($n_b \ge 0$) so that the bulk scattering matrix is diagonal
but non-diagonal boundary scattering is allowed. 
The topological charge violation at the boundary 
is incorporated following the suggestion given in 
\cite{css}. This TBA has the bulk and boundary scale dependence 
as well as the boundary parameter dependence. 

In section 3,  (R-channel) TBA 
for the massless boundary sine-Gordon model
is obtained 
as the massless limit of the massive TBA. 
It is demonstrated 
through numerical work 
that beyond a certain parameter range
($\chi \ge b^2 \pi$), 
a branch singularity crossing occurs.
Thus the original ground state TBA 
which holds within a small parameter range 
is to be modified.
We investigate on the singularity structure of TBA
and find that only one singularity of each species 
is moving on the complex plane.
This has the effect on the evaluation of the 
convolution in TBA.  From this branch 
singularity crossing information, 
we propose a modified massless TBA 
according to the suggestion 
in \cite{css,DT,DT2,lyTBA}
similarly in the bulk case \cite{BLZ}.
Section 4 is the conclusion.

\section{Massive TBA for boundary sine-Gordon} 
\noindent

The bulk sine-Gordon periodic potential allows 
a soliton with topological charge $+1$ 
and an antisoliton with the charge $-1$.  
Both of them have the same mass $M$ 
as the result of  
the charge conjugation symmetry, $\phi \to -\phi$.
The mass is given in terms of  $\mu$ \cite{mass-mu}, 
\beq
\frac{\pi \mu}{\gamma(b^2)}  =
\left[M \frac{ \sqrt{\pi} }2 \,
\frac{\Gamma(\frac1{2 -2 b^2}) } 
{ \Gamma( \frac{b^2}{2-2b^2}) } \right]^{2-2b^2}
\label{sG-m-mu}
\eeq
In addition, there are  
topologically neutral particles, breathers 
(interpreted as the soliton-antisoliton bound states).
Their masses are given as 
\beq
M_a =M m_a =2  M  \sin 
\left( \frac{\pi a}{2\lambda } \right)
\,\qquad a=1,2,\cdots, n_b\,.
\eeq
$n_b$ is the number of breather species, 
$ n_b = $ positive integer less than $ \lambda $.

The free energy density in Eq.~(\ref{massive-free-energy})
is expanded in terms of the bulk Hamiltonian eigenstates, 
{\it i.e.,} solitons, antisolitons and breathers.
These states are uniquely identified 
in terms of mass and rapidity
due to the Fermi-statistics.

The presence of the boundary 
forces two restrictions on the states. 
First, the pair creation at the boundary 
forces the rapidity paired $(\theta, -\theta)$
and therefore, one can count $\theta$ positive  
and make energy eigenvalue doubled: 
$E_{A} (\theta)  = 2 M_A  \cosh \theta $ 
where $M_A $ is the single particle mass. 

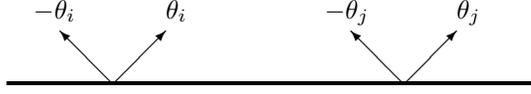
\begin{figure}[ht]
\begin{picture}(200,30) (-120,0)
\put(0,0){\line(1,0) {200}}
\put(0,0.5){\line(1,0) {200}}
\put(40,0){\vector(-1,1) {20}} \put(40,0){\vector(1,1) {20}}
\put(10,25){\footnotesize{$- \theta_i$ }}
\put(60,25){\footnotesize{$\theta_i$ }}
\put(150,0){\vector(-1,1) {20}}
\put(150,0){\vector(1,1) {20}}
\put(120,25){\footnotesize{$- \theta_j$ }}
\put(170,25){\footnotesize{$\theta_j$ }}
\end{picture}
\caption{pair creation at the boundary} 
\end{figure}

Second, at the boundary an in-coming soliton is allowed  
to be scattered away as an antisoliton and vice versa
since the soliton number is not conserved in general. 
To take care of this, soliton and antisoliton are regarded 
as a constituent of a doublet 
of identical particles \cite{css}.
When the partition function in Eq.~(\ref{massive-free-energy}) is 
written in terms of spectral density $\rho$,
$Z_{\alpha \beta} = \int [d\rho]
\exp(-RL \,f_{\alpha \beta}(L, R))$,
the spectral density should include 
not only one-particle density of 
topological particle (denoted as $\rho_0$) 
(soliton is indistinguishable  from antisoliton)
and its hole density  ($\rho_{h\, 0}$),
one-particle density of a breather ($\rho_a$),  
its hole  density ( $\rho_{h\, a}$ with $1 \le a \le n_b$)
but also should include 
two-particle density of topological particles 
($ \rho_d$) ({\it i.e.} soliton and antisoliton pair).
The densities are  summarized in the Table \ref{species}:
\begin{table}[h]
\center{ \begin{tabular}{||c|c||} \hline
\rule[-.4cm]{0cm}{1.cm}
species  &  density \\ \hline \hline
soliton or/and antisoliton   & $\rho_0\,, \rho_d\,, \rho_{h 0}$ \\ \hline
breather & $\rho_a\,, \rho_{h a}\,, \quad  a=1, \cdots, n_b$\\ 
\hline
\end{tabular}}
\caption{ particle species and the corresponding densities.} 
\label{species}
\end{table}

Then the free energy density can be written as 
\beq
R f_{\alpha \beta}(R) =  \int_0^\infty  \!\! 
d\theta \left\{  
\sum_{A=0, 1, \cdots, n_b}  \rho_A   
( 2M_A \cosh \theta - \ln \lambda_{\alpha \beta}^A (\theta) )
+ \rho_d \,( 2 M_d \cosh \theta 
- \ln \lambda^d_{\alpha \beta} ) -{\cal S_B} 
\right\}\,, \eeq 
where $M_0 =M$ and  $M_d =2 M$. 
${\cal S}_B$ is the entropy density, 
\beaq
{\cal S}_B &=& \int_0^\infty d\theta \left[ \, 
(\rho_0 + \rho_d + \rho_{h0})  \ln (\rho_0 + \rho_d  + \rho_{h0})  
-\rho_0 \ln \rho_0 -\rho_d \ln \rho_d - \rho_{h0} \ln \rho_{h0} \right.
\nn\\
&& \quad + \sum_{a=1,\cdots, n_b} \{  \left.
(\rho_a + \rho_{ha})  \ln (\rho_a +  \rho_{ha})  
-\rho_a \ln \rho_a - \rho_{ha} \ln \rho_{ha} \} \,\right] \,.
\eeaq
$\lambda_{\alpha \beta}^A  (\theta) 
= \langle B_\alpha| \theta A \rangle \, 
\langle \theta A | B_\beta  \rangle $ 
is the boundary state contribution, 
which is given in terms of the boundary scattering amplitude 
$R(u)$ with $u = -i \theta$:
\beaq
\lambda_{\alpha \beta}^a 
&=& \overline {K^a_{\alpha} (u) } K^b_{\beta} (u)
\qquad 
\textrm{ for } a=1,2, \cdots, n
\nn\\
\lambda_{\alpha \beta}^0 
&=& \overline {K_\alpha ^{++} } K_\beta^{++} 
+ \overline {K_\alpha^{+-}}  K_\beta^{+-} 
+ \overline {K_\alpha^{-+}} K_\beta^{-+} 
+ \overline {K_\alpha^{--}} K_\beta^{--} 
=\mbox{Tr}( \bar K_\alpha K_\beta )
\nn\\
\lambda_{\alpha \beta}^d 
&=& ( \overline { K_\alpha^{++} \, K_\alpha^{--}}
 -  \overline {K_\alpha^{+-} \, K_\alpha^{-+}) }
(K_\beta ^{++} K_\beta ^{--}
 -  K_\beta ^{+-} K_\beta ^{-+}) 
= \textrm{Det} ( \overline {K_\alpha} K_\beta )
\eeaq   
where $K (u) \equiv R (\pi/2 - u)$.

The boundary scattering amplitude 
(modulo CDD-type factors),
can be found in \cite{GhoZam}, which 
satisfies the boundary version of the Yang
Baxter equation, unitarity condition, 
and analyticity-crossing symmetry. 
\beaq
R(\eta ,\vartheta ,u) 
& = & 
\left( \begin{array}{cc}
R^{++}(\eta ,\vartheta ,u) &  R^{+-}(\eta ,\vartheta ,u)\\
R^{-+}Q(\eta , \vartheta , u) &  R^{--}(\eta , \vartheta , u)
\end{array}\right) 
\nn \\
 & = & \left( \begin{array}{cc}
P_{0}^{+}(\eta ,\vartheta ,u) & Q_{0}(u)\\
Q_{0}(u) & P_{0}^{-}(\eta ,\vartheta ,u)
\end{array}\right) R_{0}(u)\frac{\sigma (\eta ,u)}
{\cos (\eta )}\frac{\sigma (i\vartheta ,u)}{\cosh (\vartheta )}\, \, \, ,
\nn \\
P_{0}^{\pm }(\eta ,\vartheta ,u) 
& = & \cos (\lambda u)\cos (\eta )\cosh (\vartheta )\mp 
\sin (\lambda u)\sin (\eta )\sinh (\vartheta )
\nn \\
Q_{0}(u) & = & -\sin (\lambda u)\cos (\lambda u)\,.
\label{Rsas} 
\eeaq
Here $R_{0}$ is the boundary condition independent part,
$$
R_{0}(u)=\prod ^{\infty }_{l=1}
\left[ \frac{\Gamma (4l\lambda -\frac{2\lambda u}{\pi })
\Gamma (4\lambda (l-1)+1-\frac{2\lambda u}{\pi })}
{\Gamma ((4l-3)\lambda -\frac{2\lambda u}{\pi })
\Gamma ((4l-1)\lambda +1-\frac{2\lambda u}{\pi })}
/(u\to -u)\right] 
$$
and $\sigma (x,u)$ is the boundary condition dependence part, 
$$
\sigma (x,u)=\frac{\cos x}{\cos (x+\lambda u)}
\prod ^{\infty }_{l=1}\left[ 
\frac{\Gamma (\frac{1}{2}+\frac{x}{\pi }+(2l-1)
\lambda -\frac{\lambda u}{\pi })
\Gamma (\frac{1}{2}-\frac{x}{\pi }+(2l-1)
\lambda -\frac{\lambda u}{\pi })}
{\Gamma (\frac{1}{2}-\frac{x}{\pi }+(2l-2)
\lambda -\frac{\lambda u}{\pi })\Gamma (\frac{1}{2}+\frac{x}{\pi }+2l
\lambda -\frac{\lambda u}{\pi })}/(u\to -u)\right] \,.
$$
The scattering parameters, $\eta$ and $\vartheta$ are related   
with the action parameters, $\mu_B$ and $\varphi_0$ \cite{parameter,BPT}:
\beaq
\cos ( b^2 \eta)\, 
\cosh (b^2 \vartheta ) \, 
&=&  \mu_B \sqrt{\sin(b^2 \pi)} 
\cos  ( b \varphi_0 )/\sqrt \mu
\nn\\
\sin ( b^2 \eta)\, 
\sinh (b^2 \vartheta ) \, 
&=& \mu_B \sqrt{\sin(b^2 \pi)} 
\sin ( b\varphi_0 )/\sqrt \mu
\label{sG-parameter}
\eeaq

The boundary scattering amplitude of breathers is given as
\beq
R^{(k)}(\eta ,\vartheta ,u)
=R_{0}^{(k)}(u)\, Z^{(k)}(\eta ,u)\,
Z^{(k)}(i\vartheta ,u)\,,
\qquad 1 \le k \le n_b\,.
\eeq
$R_{0}^{(k)}$ is the boundary independent part
and $Z^{(k)}$ the boundary dependent one:
\bea
R_{0}^{(k)}(u)=\frac{\left( \frac{1}{2}\right) 
\left( \frac{k}{2\lambda }+1\right) }
{\left( \frac{k}{2\lambda }+\frac{3}{2}\right) }
\prod ^{k-1}_{l=1}\frac{\left( \frac{l}{2\lambda }\right) 
\left( \frac{l}{2\lambda }+1\right) }
{\left( \frac{l}{2\lambda }+\frac{3}{2}\right) ^{2}} \,,
\quad
Z^{(k)}(x,u)=\prod ^{k-1}_{l=0}
\frac{\left( \frac{x}{\lambda \pi }-\frac{1}{2}
+\frac{k-2l-1}{2\lambda }\right) }
{\left( \frac{x}{\lambda \pi }
+\frac{1}{2}+\frac{k-2l-1}{2\lambda }\right) }\,,
\eea
where the notation $(x)$ stands for 
\bea
 (x)=\frac{\sin \left( \frac{u}{2}+\frac{x\pi }{2}\right) }
{\sin \left( \frac{u}{2}-\frac{x\pi }{2}\right)} \,.
\eea

The hole and the particle densities are not independent each other. 
This relation is obtained from the 
bulk scattering amplitude. 
The bulk-scattering amplitude of solitons and antisolitons 
\cite{bulk-sG} are given as 
\beaq
S^{++}_{++}(u )&=& S_{--}^{--}(u)= s( u)
\nonumber \\
S^{+-}_{+-}(u)&=& S_{-+}^{-+}(u)= 
\frac{\sin (\lambda u)}{\sin (\lambda (\pi -u))}s(u)
\nonumber \\
S^{-+}_{+-}(u) &=& S_{-+}^{+-}(u)= 
\frac{\sin (\lambda \pi )}{\sin (\lambda (\pi -u))}s(u) \,
\label{soliton-s} 
\eeaq  
where $s(u)$ is given as  
\bea
s(u) =  
\prod ^{\infty }_{l=1}
\left[ \frac{\Gamma (2(l-1)\lambda-\frac{\lambda u}{\pi })
\Gamma (2l\lambda +1-\frac{\lambda u}{\pi })}
{\Gamma ((2l-1)\lambda -\frac{\lambda u}{\pi })
\Gamma ((2l-1)\lambda +1-\frac{\lambda u}{\pi })}/(u\to -u)\right] 
\eea

Due to the restriction of  $\lambda$,
the bulk scattering amplitudes are diagonal,
$ S^{-+}_{+-}(u) =0 $. 
This restriction makes our analysis 
not too much complicated \cite{sGTBA}.  
The diagonal scattering amplitude for soliton and antisoliton 
turns out to be equal up to a phase difference:
$ S^{++}_{++}(u)= (-1)^{\lambda -1} S^{+-}_{+-}(u )\,$.

The scattering amplitude of the breathers $B^a$ and $B^b$ 
with $b \le a \le n_b$ takes the form
\beq
S^{a\, b}(u)=\{a+b-1\}\{a+b-3\}\dots
\{a-b+3\}\{a-b+1\}\,,
\eeq
where the notation $\{y\}$ is defined as 
\bea
\{y\}=\frac{\left( \frac{y+1}{2\lambda }\right) \left(
\frac{y-1}{2\lambda }\right) }{\left( \frac{y+1}{2\lambda }-1\right)
\left( \frac{y-1}{2\lambda }+1\right) }
\eea
and satisfies the relations
$\{y\}\{-y\}=1\,$ and $\,\{y+2\lambda \}=\{-y\}\,$.
The scattering amplitude of the soliton (antisoliton) and breather  
$ S^{(a)}(u)  = S_{a\,+}^{a\, +}(u) = S_{a\, -}^{a\, -}(u) $
is given as  
\beq
S^{(a)}(u) =
\left\{ \begin{array}{l}
\{a-1+\lambda \}\{a-3+\lambda \}\cdots 
\{1+\lambda \}\qquad \textrm{if } a \textrm{ is even}\\
- \{a-1+\lambda \}\{a-3+\lambda \}\cdots 
\{2+\lambda \} (\frac{\lambda+1}{2\lambda})\,
(\frac{\lambda-1}{2\lambda})
\qquad \textrm{if }a\textrm{ is odd}
\end{array}\right. \,.
\eeq

Demanding the wave function periodic in $L$
we have the constraints between 
hole densities with particle densities.
For soliton states ($n_0\,, n_d\,, n_{h0}$)
we have 
\beaq
&&\exp(iL M_0 \sinh \theta_i^0 ) 
\prod_{j=1, \ne i}^N 
\Bigg\{ 
S_{00}(\theta_i^0 -\theta_j^0) 
S_{00}(\theta_i^0 +\theta_j^0) 
S_{00}(2\theta_i^0) 
S_{0d}(\theta_i^0 -\theta_j^d) 
S_{0d}(\theta_i^0 +\theta_j^d) 
\nn\\
&&\qquad\qquad 
\prod_{a=1}^{n_b} 
\bigg( S_{0a}(\theta_i^0 -\theta_j^a) 
S_{0a}(\theta_i^0 +\theta_j^a) \bigg) \Bigg \}  
=\pm e^{ 2\pi i (n_0 (\theta_i^0)
+ n_d (\theta_i^0)+n_{h0}(\theta_i^0)}\,.
\eeaq
For breathers ($n_a, n_{ah}$, $a=1,2, \cdots, n_b$):
\beaq
&&\exp(iL M_a \sinh \theta_i^a ) 
\prod_{j=1, \ne i}^N \Bigg\{ 
S_{aa}(\theta_i^a -\theta_j^a) 
S_{aa}(\theta_i^a +\theta_j^a) 
S_{aa}(2\theta_i^a) 
S_{a0}(\theta_i^a -\theta_j^0) 
S_{a0}(\theta_i^a +\theta_j^0)
\nn\\
&&\qquad\qquad\qquad\qquad
S_{ad}(\theta_i^a -\theta_j^d) 
S_{ad}(\theta_i^a +\theta_j^d) 
\prod_{b=1}^n \bigg( S_{ab}(\theta_i^a -\theta_j^b) 
S_{ab}(\theta_i^a +\theta_j^b)\bigg) 
\Bigg \}  
\nn\\
&&\qquad
=\pm e^{ 2\pi i  (n_a (\theta_i^a) +n_{ha} (\theta_i^a) ) }\,.
\eeaq
Differentiating with respect to the rapidity, 
we have the relations of hole and particle spectral densities:
\beaq
&& 
M_A \cosh \theta + 
\sum_{B=0,1, \cdots, n_b } \int_0^\infty d\theta^B
\, \rho_B(\theta^B)\, 
( \phi_{AB}(\theta^A - \theta^B) + \phi_{AB}(\theta^A + \theta^B) ) 
\nn\\
&& \qquad \qquad \qquad \qquad 
+ \int_0^\infty d\theta^d
\, \rho_d(\theta^d)\, 
( \phi_{Ad}(\theta^A - \theta^d) + \phi_{Ad}(\theta^A + \theta^d) ) 
\nn\\
&&\qquad 
= 2\pi (\rho_A(\theta ) 
+ \rho_{hA} (\theta) + \delta_{A0} \, 
\rho_d (\theta) ) \quad \textrm{ for } A=0, 1, \cdots, n_b, d,
\eeaq
where 
$\rho_A (\theta) = \frac1L \frac {d n_A (\theta)} {d\theta}$,
$\phi_{AB}(\theta) 
= -i \frac{d \ln S_{AB} (\theta)}{d \theta}$
and 
$\phi_{Ad} (\theta) = 2 \phi_{A0}(\theta)$.

Introducing pseudo energies,  $\epsilon$ 
\bea
e^{-\epsilon_a} =\frac{\rho_a}{\rho_{ha}} 
\quad \textrm{ for } a =1, \cdots, n_b\,, \quad 
e^{-\epsilon_0} = \frac{\rho_0}{\rho_{h0}}  \,,\quad
e^{-\epsilon_d} = \frac{\rho_d}{\rho_{h0}}   \,,
\eea 
and minimizing $f_{\alpha\beta}(R)$ we have 
the massive TBA:
\beaq
\epsilon_A (\theta) 
&=& d_A (\theta)- \ln \lambda_{\alpha \beta}^A 
-\frac1{2\pi}\sum_{B=0,1,\cdots, n_b} 
 \int_{-\infty}^\infty d \theta' 
\phi_{AB}(\theta -\theta') L_B(\theta') \,,
\nn\\
\epsilon_d (\theta) 
&=&  2(\epsilon_0  +  \ln \lambda_{\alpha\beta}^0 ) 
- \ln \lambda_{\alpha\beta}^d  \,,
\eeaq
where $d_A (\theta) =2 M_A R \cosh \theta\,,\,\,$
$L_0 = \ln (1 + e^{- \epsilon_0} +e^{-\epsilon_d})\,,\,\,$
and 
$L_a = \ln (1 + e^{- \epsilon_a})\,$ ( $a =1, \cdots, n_b $).

Shifting the pseudo energy 
$\epsilon_A \to \epsilon_A - \ln \lambda_A$, 
we have $\epsilon_d = 2 \epsilon_0$ and 
the boundary TBA is put equivalently as
\beq
\epsilon_A (\theta)
= d_A (\theta) -\frac1{2\pi}\sum_B 
 \int_{-\infty}^\infty d \theta' \,
\phi_{A\, B}(\theta -\theta') \,
\widetilde {L_{B}}(\theta') 
\label{bsG-TBA}
\eeq
where $A=0,1,\cdots, n_b$ and 
\bea
\widetilde {L_0} &\equiv & \ln (1 + \lambda^0_{\alpha\beta}\,
e^{- \epsilon_0} + \lambda^d_{\alpha\beta} \,e^{-2\epsilon_0})\,
\nn\\
\widetilde {L_a} &\equiv & 
\ln (1 + \lambda^a_{\alpha \beta} \, e^{- \epsilon_a}) 
\qquad\qquad 
\textrm{ for } a =1, \cdots, n_b\,.
\eea
In terms of this TBA, the free energy has the form,
\beq
E(R)= 
R\,f(R) = R \mathcal{E}_{\rm bulk} 
+ \mathcal{E}_{\rm boundary} 
- \frac{\pi}{24 R} c_{\rm eff} 
\label{sG-energy}
\eeq
where $\mathcal{E}_{\rm bulk}
= - \frac14 M^2 \tan \frac{\pi}{ 2\lambda}$ 
is the bulk energy density,
$\mathcal{E}_{\rm boundary} $ is the boundary energy 
obtained by Al. Zamolodchikov \cite{parameter}, 
whose details can be found in \cite{BPT}
\bea
\mathcal{E}_{\rm boundary} (\eta, \vartheta) 
= - \frac M {2 \cos (\pi/2\lambda)}
\Bigg(\cos (\frac \eta \lambda) 
+ \cosh (\frac \vartheta \lambda) 
-\frac12 \cos (\frac \pi{2 \lambda}) 
+ \frac 12 \sin (\frac \pi {2 \lambda}) - 
\frac12 \Bigg)\,,
\eea
and $c_{\rm eff}$  is the effective central charge
\bea
c_{\rm eff} (RM) = \frac{6RM}{\pi^2 } 
\int_{-\infty}^\infty d \theta \,
\sum_{A=0,1,\cdots, n_b} m_A \,\cosh \theta \,  
\widetilde {L_A}(\theta)\,.
\eea

One may also put this TBA in a reduced form
as in a bulk TBA \cite{ade} for $n_b \ge 1$.
To do this, we include the doublet formally by 
extending the whole index into 
$A' =0,1,\cdots, n_b, + $
\beq
\epsilon_{A'}(\theta)
= d_{A'}(\theta) -\frac1{2\pi}\sum_{ B'=0,1,\cdots, n_b, + }
 \int_{-\infty}^\infty d \theta' 
\phi_{A'\, B'}(\theta -\theta')
\widetilde {L_{B'}}(\theta')
\eeq
where $+$ denotes the doublet,
$\epsilon_+ \equiv \epsilon_d /2 $,  
$\, m_+ \equiv m_0$, and $\, \widetilde {L_+} \equiv 0\,$.
Then using an identity of the fourier transformed form of the kernel,
$\tilde \varphi_{A'B'} (k) 
= \int_{-\infty}^{\infty}d\theta  \,
\phi_{A'B'}(\theta)   \,e^{ik\theta} $,
\beq
\left( \delta_{A'B'} - \frac1{2\pi} \tilde \varphi_{A'B'} (k) \right)^{-1}
=  \delta_{A'B'} - \frac1{2 \cosh(k\pi /h) } \mathcal I_{A'\,B'}  
\label{kernelrelation}
\eeq
where $h=2\lambda = 2n_b +2 $ and 
$\mathcal I_{A'\,B'}$  is the incidence matrix of $D_{n_b+2}$,
\cite{FSW,ade,KM} 
($\mathcal I_{A'\,B'}=1 $ if  ${A'\,B'}$ 
is directly connected, 0 otherwise),
\vskip 0.5cm
\noindent 
\centerline
{\hbox{\rlap{\raise28pt\hbox{$\hskip6.5cm\bigcirc\hskip.25cm 0$}}
\rlap{\lower27pt\hbox{$\hskip6.4cm\bigcirc\hskip.3cm +$}}
\rlap{\raise15pt\hbox{$\hskip6.1cm\Big/$}}
\rlap{\lower14pt\hbox{$\hskip6.0cm\Big\backslash$}}
\rlap{\raise15pt\hbox{$1\hskip1.1cm 2
	\hskip1.5cm s\hskip 0.7cm \ n_b-1$}}
$\bigcirc$------$\bigcirc$-- -- --
--$\bigcirc$-- -- --$\bigcirc$------$\bigcirc$\hskip.1cm 
$\ n_b$ }} \\
\noindent
we can put the TBA  in a reduced form when  $n_b \ge 1$,
\beq
\epsilon_{A'}(\theta)
= d_{A'}(\theta) 
+ \sum_{B'=0,1,\cdots, n_b, + } \mathcal I_{A'\,B'} \,
\int_{-\infty}^{\infty} d\theta'\,
K(\theta-\theta') (\widetilde{L_{B'}} 
+\epsilon_{B'} -d_{B'})(\theta') \,,
\label{r-bsG-TBA}
\eeq
where $K(\theta)$ is the reduced kernel,
\bea
K(\theta) = \frac \lambda{2\pi \cosh(\lambda\, \theta)}
\eea
This reduced TBA allows one to construct Y-system equations with 
$Y(\theta) = e^{\epsilon(\theta)}$, 
\beaq
Y_0(\theta + \frac{i\pi}h ) \,
Y_0(\theta - \frac{i\pi}h ) \,
&=& [ \lambda_{n_b}(\theta) + Y_{n_b}(\theta)] 
\nonumber\\
Y_{n_b}(\theta + \frac{i\pi}h )\, 
Y_{n_b}(\theta - \frac{i\pi}h ) 
&=& [ \lambda_d(\theta) 
+ \lambda_0(\theta)\,Y_0(\theta)
 + Y_0(\theta)^2 ]  
[ \lambda_{n_b-1}(\theta) + Y_{n_b-1}(\theta)] 
\nonumber\\
Y_a(\theta + \frac{i\pi}h ) \,
Y_a(\theta - \frac{i\pi}h ) 
&=& \prod_b [ \lambda_b(\theta) 
+ Y_b(\theta)]^{ \mathcal I_{ab} } 
\qquad \mathrm{ for }\,\,\, a=1,2,\cdots, n_b-1
\nonumber\\
Y_+(\theta) &=& Y_0(\theta)\,.
\label{Y-system}
\eeaq


\section{Massless TBA for boundary sine-Gordon model} 
\noindent 

The massive TBA shows the scale dependence as well as 
the parametric dependence.  
Since we are particularly interested 
in the parametric dependence of the boundary TBA
on $\phi_0$, we will look at the massless limit
of the TBA in this paper.
 
The massless TBA corresponding to bSG action 
Eq.~(\ref{massless-sG-action}) is obtained 
by taking the limit $\mu \to 0$ 
of the massive TBA in Eq.~(\ref{bsG-TBA}, \ref{r-bsG-TBA}). 
Even though the soliton mass $M$ vanishes,
one may introduce a finite renormalized mass scale $M_R$  
as $ M_R = (M/2) e^{\theta_0} $
if a large parameter $\theta_0$ is defined as 
\bea
e^{-\theta_0} =  (C_0\, \mu)^{\frac {\lambda +1}{2 \lambda}}/M_R \,,
\qquad 
C_0= \frac{\pi}{\gamma(b^2)} 
\left(
\frac{\Gamma (\frac{b^2}{2-2 b^2}) }
{\sqrt{\pi} \Gamma (\frac{1}{2-2 b^2})  }
\right)^{2\lambda/(\lambda+1)}\,.
\eea
In this limit, the rapidity is rescaled 
into renormalized one, $\theta_R $
as $\theta = \theta_R + \theta_0$.
The boundary scattering parameter, $\vartheta $ 
is also rescaled as $\vartheta_R $ 
maintaining a relation
$ \vartheta - \lambda \theta  = \vartheta_R - \lambda \theta_R  $.
Then $\vartheta_R $ is written in terms of the action parameters,  
\beq 
(m_R)^{b^2 \lambda} \,
e^{b^2 \vartheta_R  } 
=  2\mu_B \sqrt{C_0 \,\sin (b^2 \pi)}  \,.
\eeq
On the other hand, $\eta$ 
is not rescaled but is identified as 
\beq
b^2  \eta = \left\{
\begin{array}{ll}
b \phi_0  \qquad  & {\rm for} 
\qquad 0 \leq b \phi_0  \leq \pi/2 \,,
\\
\pi - b \phi_0  \qquad &{\rm  for}
\qquad \pi/2 \leq b \phi_0  \leq \pi \,.
\end{array}
\right.
\label{eta-parameter}
\eeq
This identification is justified 
from the numerical analysis later on.
 In terms of this rescaled parameter 
(we omit hereafter the subscript $R$ 
standing for the renormalized one)
the solitonic boundary scattering amplitudes are given as 
\beq
K(\eta ,\vartheta ,u) \to
\left( \begin{array}{cc}
 e^{-i \eta} e^{\frac12( \vartheta +i\lambda \tilde u)}&
-i e^{-\frac12( \vartheta +i\lambda \tilde u)} \\
-i e^{-\frac12( \vartheta +i\lambda \tilde u)} &  
e^{i \eta} e^{\frac12( \vartheta +i\lambda \tilde u)}
\end{array}\right) 
e^{ -i 3 \pi \lambda/ 4 } \, k(u) 
\eeq
where $\tilde u = \pi/2 -u$ and 
$ k( u)^{-1}  =\prod_{q=0}^{n_b} {2 \cos \left(\frac{\pi}{2 \lambda}
\left(\frac12 + q+ 
\frac{i\vartheta -  \lambda \tilde u }{\pi} \right)\right)}\,$,
and the breather boundary reflection amplitudes are given as 
\beq
K^{(k)}(\vartheta, \eta, u)  \to  \prod_{l=0}^{k-1}
\frac { \sin(\frac12(\tilde u - 
\frac{i\vartheta}\lambda 
- \frac\pi2 - \pi \frac{k-2l -1}{2\lambda})) }
{\sin(\frac12(\tilde u - \frac{i\vartheta}\lambda
 + \frac\pi2 - \pi \frac{k-2l -1}{2\lambda}))  }
\eeq
The boundary $ \lambda_{\alpha\beta} $ is given as
\beaq
\lambda^0_{\alpha\beta} &=& 2 \left\{ \cos \eta\, 
e^{(\frac{\vartheta_\alpha +\vartheta_\beta}2 - \lambda \theta)} 
+ e^{-(\frac{\vartheta_\alpha 
+\vartheta_\beta}2 - \lambda \theta)} 
\right\} \,\overline{ k_\alpha (u)} \,k_\beta (u) \,,
\nn\\
\lambda^d_{\alpha\beta} &=&
\left\{ e^{(\vartheta_\alpha 
-\lambda \theta - i\pi\lambda/2 ) } 
+e^{-(\vartheta_\alpha -\lambda \theta - i\pi\lambda/2 ) }
\right\}\,
\nn\\
&&\qquad 
\left\{ 
e^{(\vartheta_\beta -\lambda \theta +i\pi\lambda/2 )} 
+ e^{-(\vartheta_\beta -\lambda \theta + i\pi\lambda/2 ) } 
\right\} \, 
\left(\overline{ k_\alpha(u)} \, k_\beta(u) \right)^2
\label{massless-lambda}
\eeaq

From this massless scattering data 
the massless TBA is given as the form  
\beq
\epsilon_A(\theta) =
D_A (\theta) -\frac1{2\pi}\sum_{ B } 
 \int_{-\infty}^\infty d \theta' \,
\phi_{A\, B}(\theta -\theta') \,
\widetilde {L_{B}}(\theta')
\label{m0sG-TBA}
\eeq
where $D_{A'} = 2 m_{A'} r e^\theta $ with $r=MR$ 
and the soliton mass $M$ in $M_{A'}$ 
is replaced by the renormalized mass $M_R$.
The reduced form of TBA can be similarly  obtained.
TBA Eq.~(\ref{m0sG-TBA}), however, turns out to be better suited  
for the numerical analysis of various range of parameters. 
The free energy is given as 
\beq
R\, f(R) =\mathcal{E}_{\rm boundary} 
- \frac \pi{24 R} c_{\rm eff}
\label{m0sG-energy}
\eeq
where the bulk energy density vanishes,
and the boundary energy and effective central charge are given as 
\bea
\mathcal{E}_{\rm boundary} 
&=& - \frac M{2 \cos (\pi/2\lambda)} e^{\vartheta/\lambda}\\
c_{\rm eff} &=&  \frac{12 \,r}{\pi^2} \int_{-\infty}^\infty d \theta
\,\sum_{A=0,1,\cdots, n_b} m_{A }  \,e^\theta \, \widetilde {L_A}(\theta)\,.
\eea 

Let us investigate the parametric dependence of 
the ground state energy 
Eq.~(\ref{m0sG-energy}) using TBA Eq.~(\ref{m0sG-TBA}).
For the details of analysis, 
we will restrict our selves to 
$\lambda=1$ and $\lambda=2$ cases with 
the symmetric boundaries 
($\vartheta_\alpha = \vartheta_\beta = \vartheta$). 

When $\lambda =1 ( n_b=0 )$,
the boundary contribution is given 
explicitly as, 
\bea
\lambda^0_{\alpha\beta} =
\frac{
2\left[\cos \eta\, 
e^{(\frac{\vartheta_\alpha +\vartheta_\beta}2 -  \theta)} 
+ e^{-(\frac{\vartheta_\alpha +\vartheta_\beta}2 - \theta)} 
\right]}
{4 \cosh (\frac{\vartheta_\alpha - \theta}2 )
\cosh (\frac{\vartheta_\beta - \theta}2 )} 
\,,\qquad
\lambda^d_{\alpha\beta} =
\tanh (\frac{\vartheta_\alpha -\theta}{2} )
\tanh (\frac{\vartheta_\beta -\theta}{2})\,.
\eea

\begin{figure}[h]
\begin{minipage}[t]{7.5cm}
{\scalebox{0.45} {\includegraphics{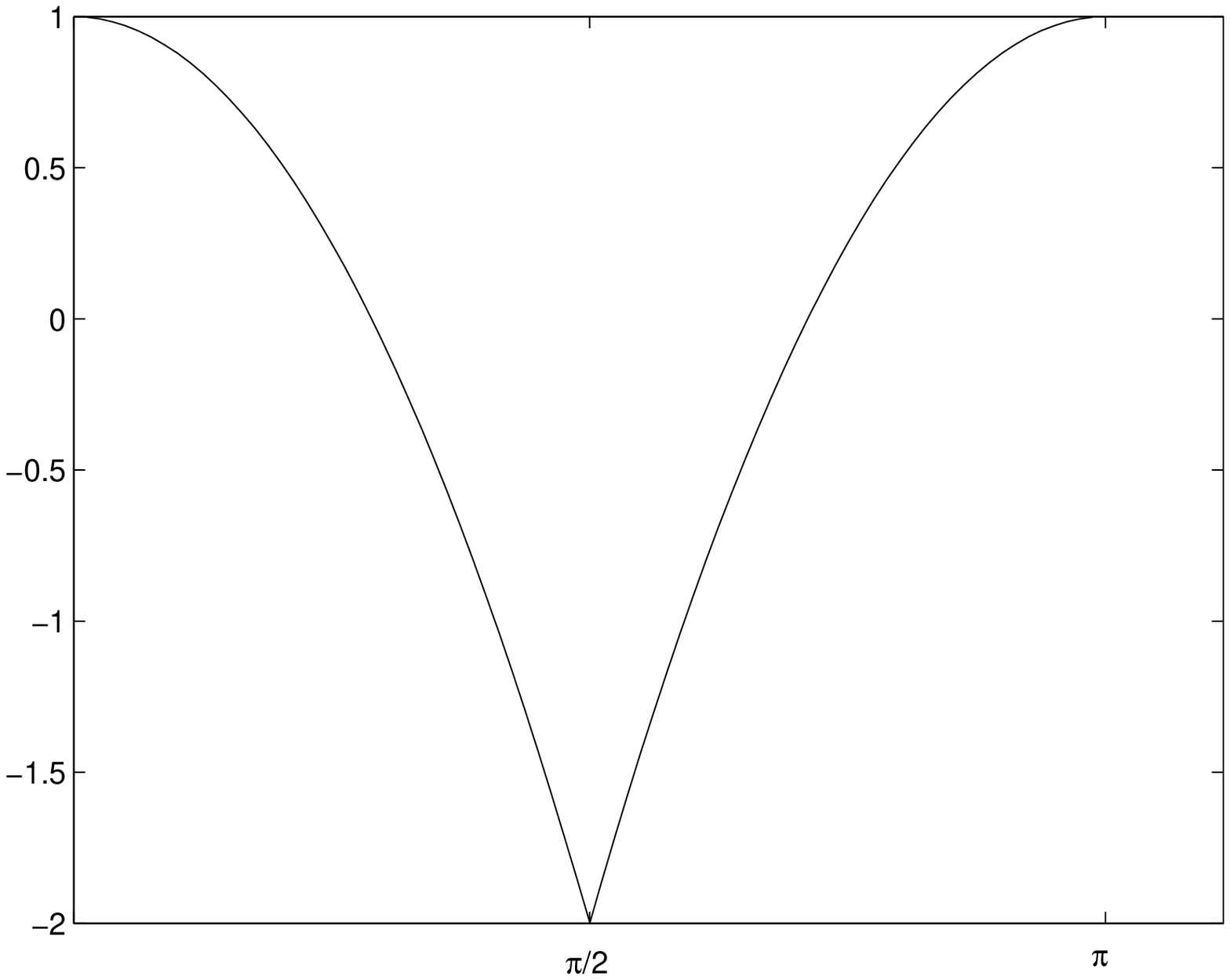}}}
\caption{$c_{\rm eff}$ vs. $\!\chi\,$  
when $\lambda=1$ and $\vartheta = 10 $ 
before modifing the massless TBA.}  
\label{fig4} 
\end{minipage}
\ $\qquad$ \
\begin{minipage}[t]{7.5cm}
{\scalebox{0.45} {\includegraphics{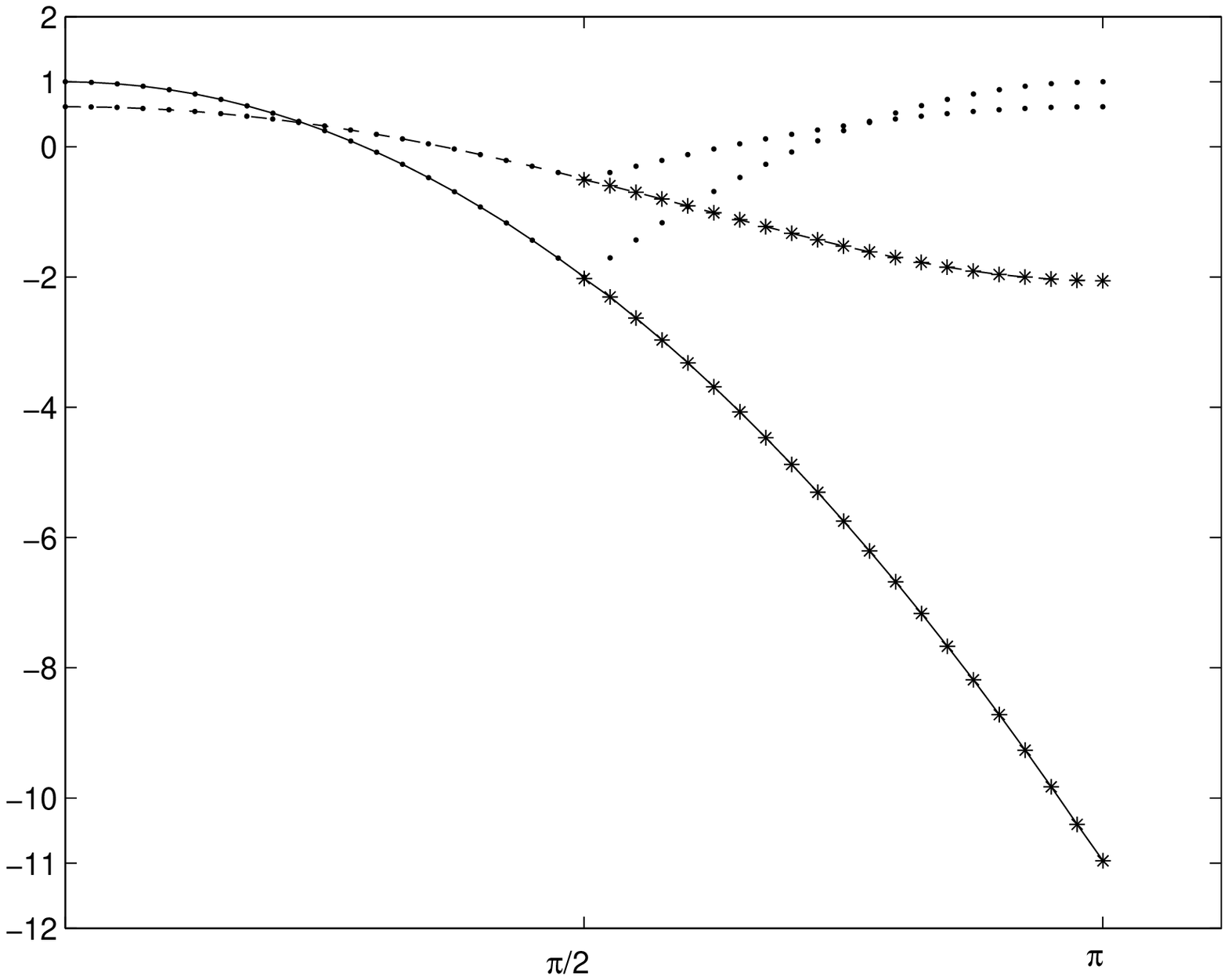}}}
\caption{$c_{\rm eff}$ vs. $\!\chi\,$ when $\lambda=1$. 
Modified TBA shows the correct behavior (in stars)
for $\vartheta=10$ (sold) and $\vartheta=0$ (dashed).}  
\label{fig5}
\end{minipage}
\end{figure}

TBA is trivial since the kernel $\phi(\theta)=0$:
$\epsilon_0 (\theta) = 2 m_0 r e^\theta$. 
The effective central charge 
is obtained numerically and is plotted 
$c_{\rm eff} $ v.s. $\chi$ in Fig.~\ref{fig4} where
$\chi = b(\phi_0^{(2)} - \phi_0^{(1)})$.
(One may put $\phi_0^{(1)}=0$ 
using the freedom of field translation).
We note that 
$c_{\rm eff}$ is  
$\pi$-periodic in $\chi$. 
This is because the boundary fugacity  
$\lambda_{\alpha\beta}^0$ in 
Eq.~(\ref{massless-lambda}) is $2\pi$-periodic 
in $\eta =\eta_1 -\eta_2$ and  
$\lambda_{\alpha\beta}$ is $\eta$ independent.  
Generally, $c_{\rm eff}$ will be 
$2\pi b^2$- periodic in $\chi$. 

However, the periodicity of the energy in $\chi$ 
is not acceptable as pointed out in \cite{css}.
When $\chi >  \pi/2$, the boundary term in the Lagrangian 
effectively changes the relative sign; one can 
equivalently put $\mu^{(1)}_B \to -\mu^{(1)}_B$ 
and $\mu^{(2)}_B \to \mu^{(2)}_B$ while $\chi \to  \pi - \chi $.
This relative sign change of the boundary term 
should be reflected in the $c_{\rm eff}$ value. 
The same problem
was also observed in boundary Lee-Yang model \cite{lyTBA}. 

To cure this disease it has been proposed in \cite{css,lyTBA}
that a certain excited state contribution should be properly taken care of.  
According to this proposal, the  $c_{\rm eff}$ is recalculated 
and is presented in Fig.~\ref{fig5} when $\lambda=1$.
In the analysis two things are considered for $\chi>\pi/2$: 
First, one needs to find the zeroes of $e^{\tilde L_0}$ 
in terms of 
the complex rapidity $\tilde \theta$ following \cite{DT}
with the parameter identification in Eq.~(\ref{eta-parameter}).
\beq 
1 + \lambda^0_{\alpha\beta}\, 
e^{- \epsilon_0 ( \tilde \theta )} 
+ \lambda^d_{\alpha\beta} 
\,e^{-2\epsilon_0 ( \tilde \theta )} =0\,.
\label{zer0}
\eeq
Eq.~(\ref{zer0}) is solved numerically. 
It turns out that there are 
infinite number of solutions for
$ \tilde \theta = \pm i \pi/2 + \theta_p$
with $\theta_p$ real
and therefore, the pseudo energy is given as 
$\epsilon_0 ( \tilde \theta ) = \pm i 2 m_0 r e^{\theta_p}$.

\begin{figure}[h]
{\scalebox{0.95} {\includegraphics{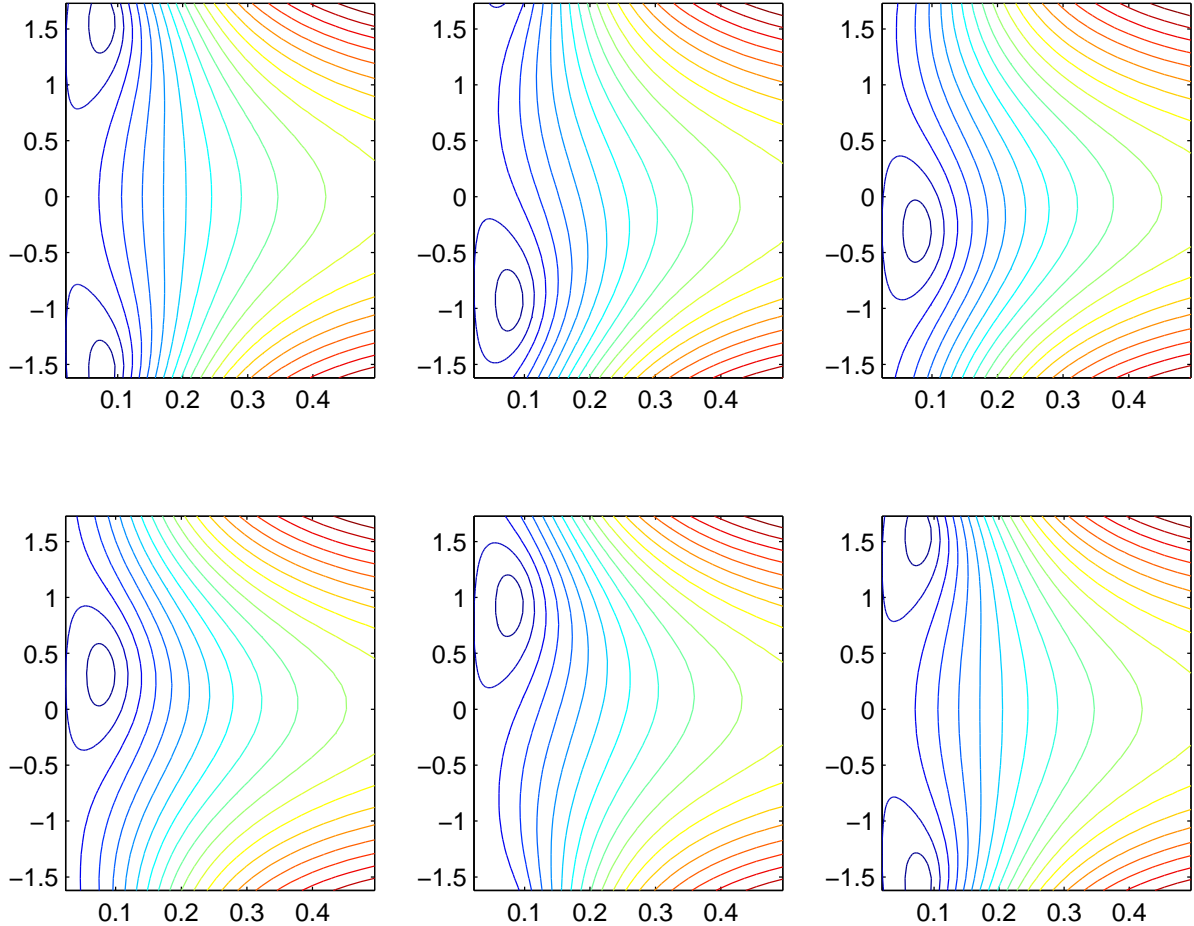}}}\
\caption{Singularity crossing for $\lambda=1$ case: 
Contour plot of $ 1/(1+ |t_0|)$
is presented in the complex $\theta$ plane.
The vertical axis stands for 
$-\frac{\pi}2 \le \rm{Im}(\theta)\le \frac{\pi}2$
and the horizontal one for $0< e^{\rm {Re}(\theta)}<0.5$.
From the top left to the bottom right, 
each figure is reproduced for 
$\chi=\pi/2-\frac{\pi}{20} e^{i \pi k }$ 
with $k=0,1/5, 2/5, 3/5, 4/5, 1$ for $\vartheta=0$.}
\label{crossing-1}
\end{figure}

Second, one needs to take account 
this branch singularity effect 
into the free energy.
To understand which solution has the effect 
on the free energy for $\chi >\pi/2 $,
we just complexify the value of $\chi$ (equivalently $\phi_0$) 
around $\pi/2$ (cusp point) 
as $\chi \to \pi/2 - (\frac\pi{20}) e^{i \pi k}$
with $k$ varying from $0 \to 1$
and use the TBA Eq.~(\ref{m0sG-TBA})
allowing $\theta$ complex but $\theta'$ real
to trace the movement of singularities.
With the notation 
$t_0 \equiv e^{\tilde L_0} = 
1 + \lambda^0_{\alpha\beta}\, e^{- \epsilon_0 } 
+ \lambda^d_{\alpha\beta} 
\,e^{-2\epsilon_0} $, 
the result is 
given in Fig.~\ref{crossing-1} for $\vartheta=$.
We choose the case of 
$\vartheta=0$ since this does not 
correspond to the conformal boundary condition,
neither Dirichlet not Neumann condition.
(Note that the singularity position is contour plotted 
using $ 1/(1+ |t_0|)$ rather than $ 1/|t_0|$ 
to renormalize the height).

\begin{figure}[h]
{\scalebox{0.95}{\includegraphics{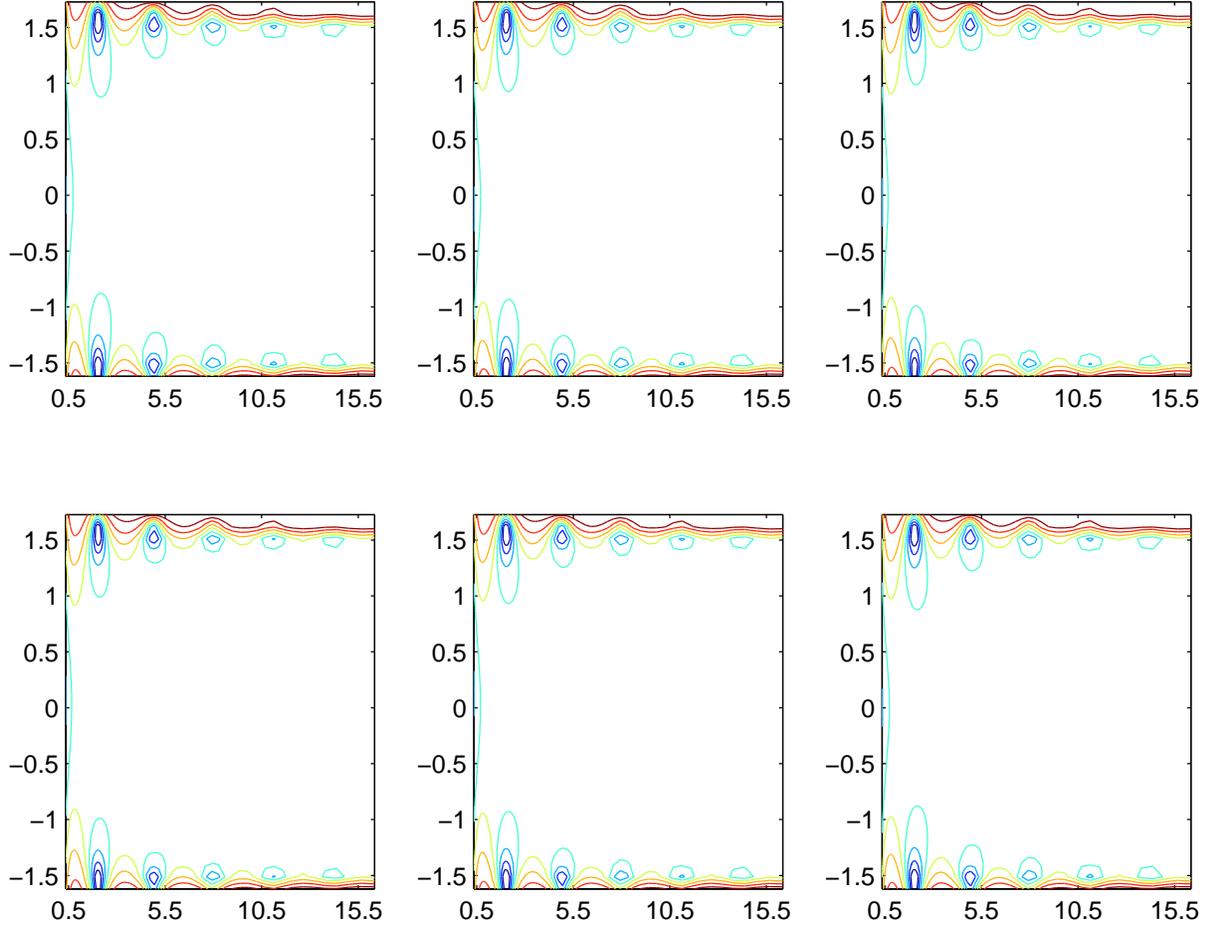}}}\
\caption{Other singularities for $\lambda=1$ case: 
Contour plot of $ 1/(1+ |t_0|)$
is presented in the complex $\theta$ plane.
The vertical axis stands for 
$-\frac{\pi}2 \le \rm{Im}(\theta)\le \frac{\pi}2$
and the horizontal for $0.5  \le e^{\rm {Re}(\theta)} \le 16 $.
From the top left to the bottom right, 
each figure is reproduced for 
$\chi=\pi/2-\frac{\pi}{20} e^{i \pi k }$ 
with $k=0,1/5, 2/5, 3/5, 4/5, 1$ for $\vartheta=0$.}
\label{nocrossing-1}
\end{figure}

When $\chi =$ real, 
singularities lie on 
the imaginary value Im($\tilde \theta) =\pm i \pi/2$.
Only one of the singularities 
(the left-most one with the smallest value of $\theta_p$)
crosses the real rapidity line
during the complexified changes of $\chi$
from the  negative imaginary value $-i\pi/2$ 
to the positive imaginary plane
and finally sits at the positive imaginary 
value $i\pi/2$.
(When  $k$ varies from $0$ to $- 1$,  
the left-most singularity  
at $i\pi/2$ is moving down and crosses the real axis).
The other signularities remains in the 
imaginary axis with Im($\theta$) = $\pm \pi/2$.
This is seen in 
Fig.~\ref{nocrossing-1}.
Similar behavior is demonstrated in 
scaling Lee-Yang model \cite{DT,DT2,BLZ} 
and used in boundary SLYM \cite{lyTBA}.
(Crossing singularities 
are paired for massive theories).

Due to this singularity crossing, 
the rapidity integration in the free energy
has to detour the branch singularity.
As the singularity crosses the real axis
and moves up to Im($\tilde \theta) =  \pi/2$
and one has to evaluate the 
contour integration around the 
singularity $\tilde \theta = i \pi/2 + \theta_p$.
The result is given as 
\beq
-\frac \pi{24 r} c_{\rm eff} = -im_0 \,e^{\tilde \theta}
- \frac1{2\pi} -\!\!\!\!\!\!\int _{-\infty}^\infty d \theta
\, m_0 \,e^\theta \, \tilde L_0(\theta)\,,
\eeq
where $ -\!\!\!\!\!\int $ represents 
the principal value of the integration.  
The the first term in the RHS 
is the singularity crossing contribution 
and the second one is the real axis contribution.
(Here we adopt the convention of the singularity crossing 
from the negative imaginary value to the positive imaginary one.
One may equivalently consider the opposite singularity crossing 
and detour the 
integration down to the negative imaginary complex plane
and get the same result. 
Note that if $\tilde \theta $ is a solution, 
so is $\tilde \theta^*$).

We generalize this idea into the case 
$\lambda >1 $. 
To understand the singularity crossing, 
we need to know
the singularity structure in the comlex $\theta$ plane.
Note that the TBA Eq.~(\ref{m0sG-TBA}) is 
singularity-free on the rapidity domain with 
$-\pi/h < \rm{Im}(\theta) < \pi/h $
when $\chi$ is real.
One can analytically continue the TBA 
into other domain of $\theta$ 
using the relation 
of Y-system Eq.~(\ref{Y-system}).

When $\lambda=2$ 
the boundary $\lambda_{\alpha \beta}$ is given as
\beaq
\lambda^0_{\alpha\beta} &=& 
\frac{ 2 \left[\cos \eta\, 
e^{(\frac{\vartheta_\alpha +\vartheta_\beta}2 -2  \theta)} 
+ e^{-(\frac{\vartheta_\alpha +\vartheta_\beta}2 - 2\theta)} \right]}
{\left[
4 \cosh (\frac{\vartheta_\alpha - 2\theta}4
 + \frac{i\pi}8)
\cosh (\frac{\vartheta_\alpha - 2\theta}4
-\frac{i\pi}8 )
\right] \left[
4 \cosh (\frac{\vartheta_\beta - 2\theta}4
 + \frac{i\pi}8)
\cosh (\frac{\vartheta_\beta - 2\theta}4
-\frac{i\pi}8 ) \right]} \,,
\nn\\
\lambda^d_{\alpha\beta} &=&
\frac{
4 \cosh (\vartheta_\alpha - 2\theta )
\cosh (\vartheta_\beta - 2\theta )}
{\left( 
4 \cosh (\frac{\vartheta_\alpha -2\theta}{4} + \frac{i\pi}8)  \cosh (\frac{\vartheta_\alpha -2\theta}{4} - \frac{i\pi}8) 
\right)^2
\left(
4 \cosh (\frac{\vartheta_\beta -2\theta}{4} + \frac{i\pi}8) 
\cosh (\frac{\vartheta_\beta -2\theta}{4} - \frac{i\pi}8) \right)^2 }  \,, \nn\\ 
\lambda^1_{\alpha \beta} &=& 
\tanh ( \theta/2  - \vartheta_\alpha /4 )\,
\tanh ( \theta/2  - \vartheta_\beta /4 )\,,
\eeaq 
where $\eta$ parameter is identified according to 
Eq.~(\ref{eta-parameter}) and the kernel is given as 
\bea
\phi_{00}(\theta) = \frac12 \phi_{11}(\theta) 
=- \frac 1{\cosh \theta} \,,
\qquad 
\phi_{01}(\theta) = 
- \frac {2\sqrt2 \cosh(\theta)}{\cosh(2\theta)}\,.
\eea

\begin{figure}[h]
\center{
{\scalebox{0.65}{\includegraphics{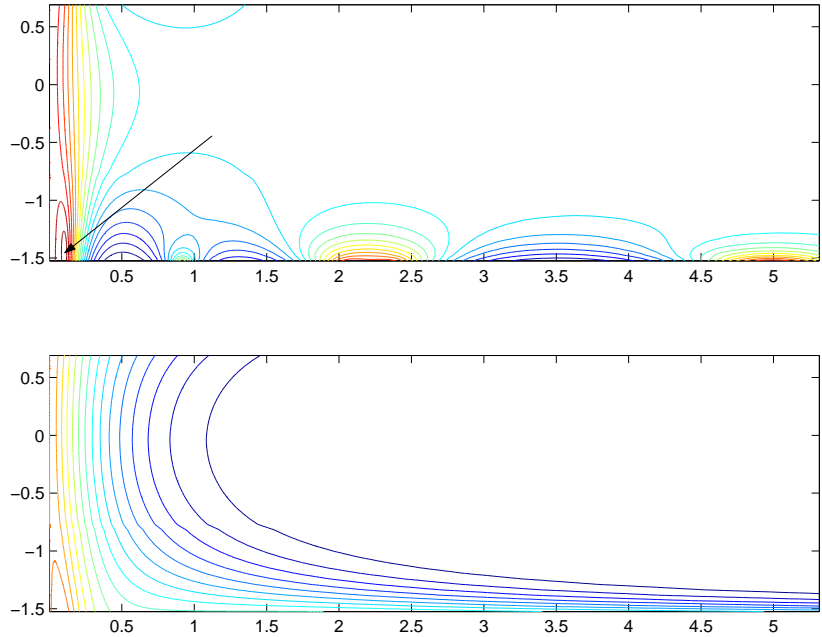}}}}
\caption{Singularity structure of TBA
in the complex $\theta$ plane when $\lambda=2$ 
and $\chi =\pi/3 -\pi/30$:
Contour plot of $1/(1+ |t_0|)$ (up) and 
$1/(1+|z_1|)$ (down).
The vertical axis stands for 
$-\frac{\pi}2 \le \rm{Im}(\theta)\le \frac{\pi}4$
and the horizontal one for 
$0 \le e^{\rm {Re} (\theta)} \le 6$.
The $t_0$ zero with the arrow attached is moving.}
\label{2-singluar}
\end{figure}

$c_{\rm eff}$ is $\pi$-periodic in $\chi$ and 
has the cusp at $\pi/3$. 
(See Fig.~\ref{2-ceff} below). 
Fig.~\ref{2-singluar} is the 
numerical study of the singularity 
in the domain with Im($\theta)\le \pi/2$
at real $\chi =\pi/3 -\pi/30$ below
the cusp point. 
Here $t_0 \equiv e^{\widetilde{L_0}}$.
In the figure $z_1\equiv e^{\epsilon_1}$
and $1/(1+|t_0|)$ and  $1/(1+|z_1|)$
are plotted  after normalizing $1/|t_0|$ and  $1/|z_1|$.

\begin{figure}[h]
\begin{minipage}[t]{7.5cm}
{\scalebox{0.45} {\includegraphics{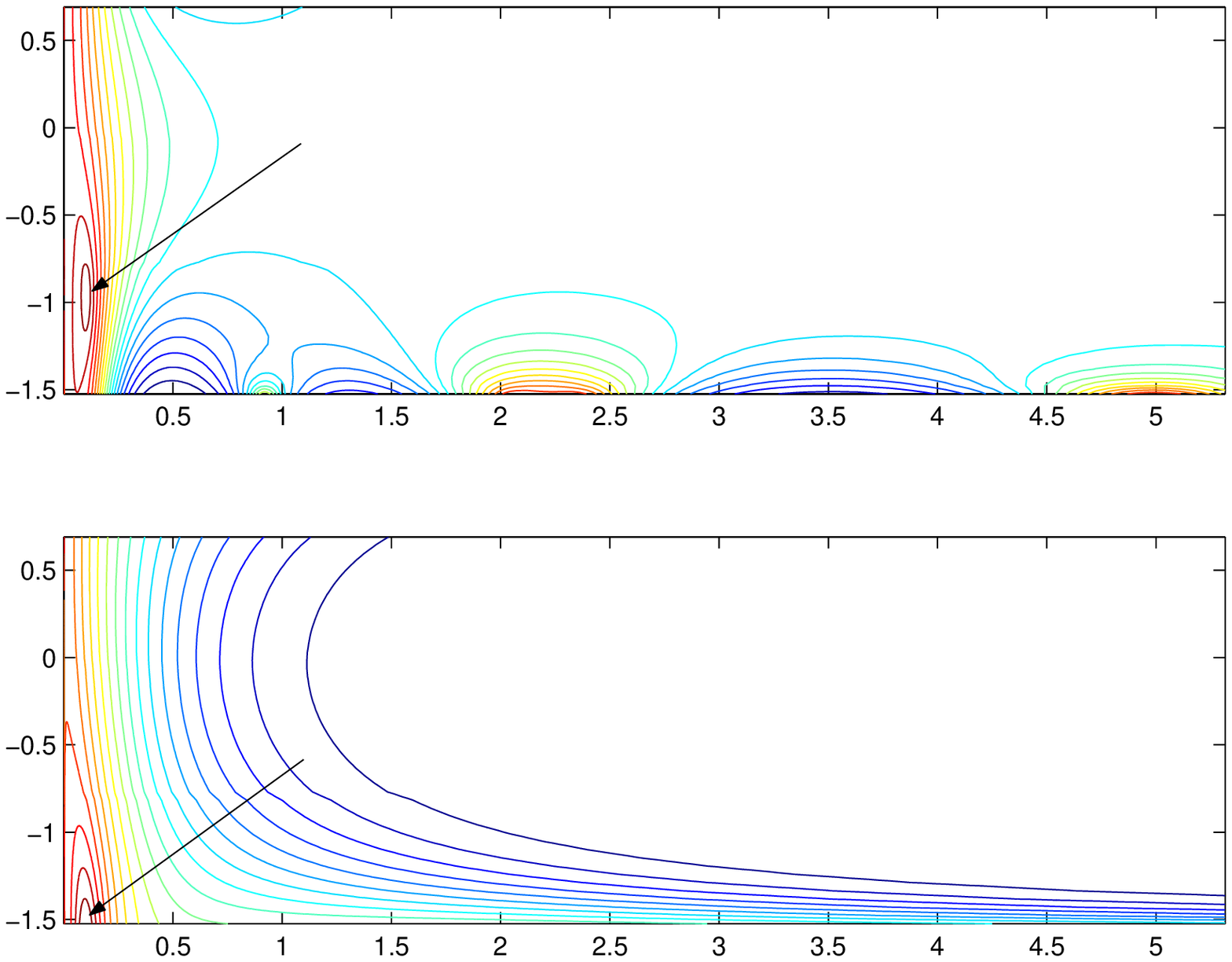}}}
\end{minipage}
\ $\qquad$ \
\begin{minipage}[t]{7.5cm}
{\scalebox{0.45} {\includegraphics{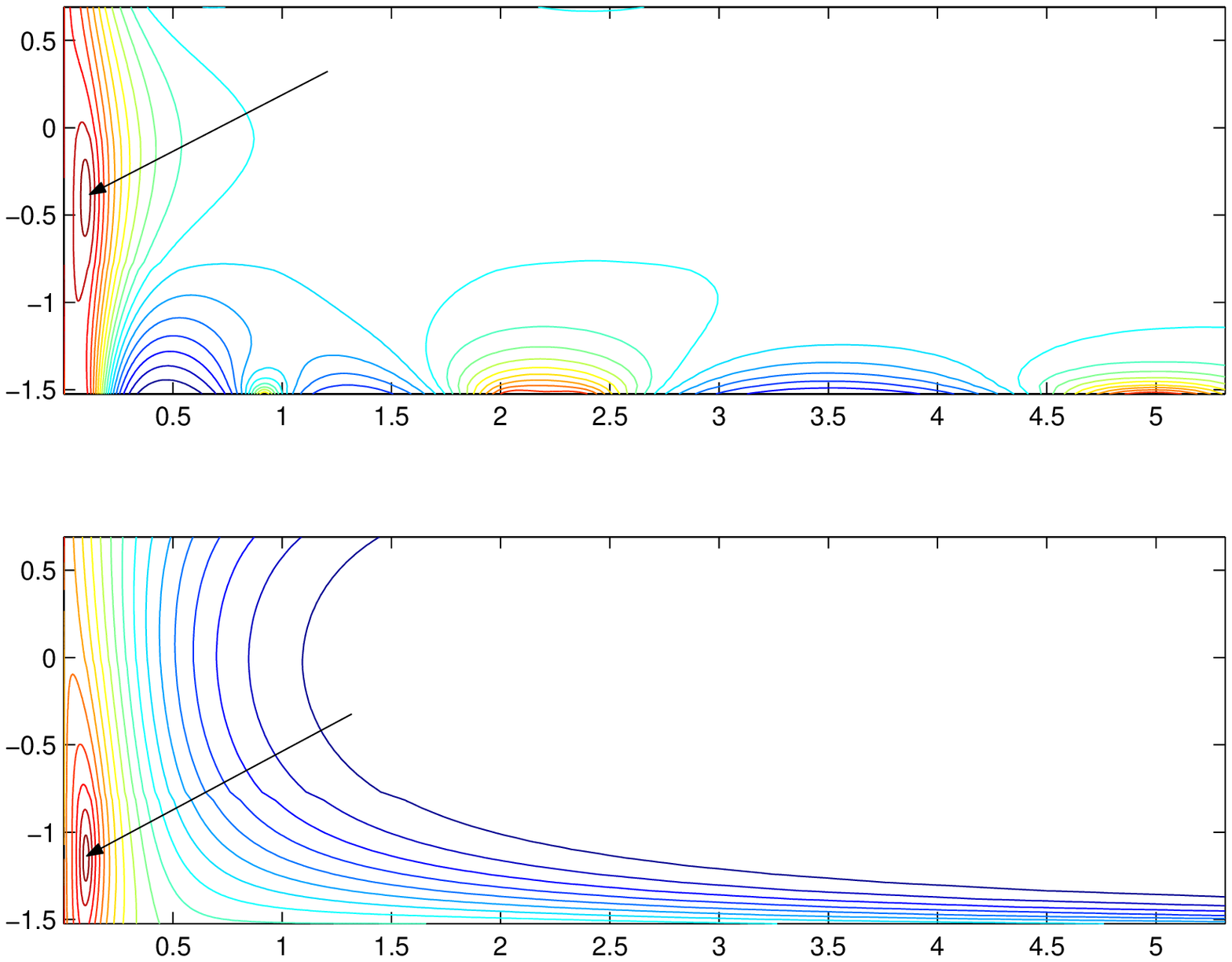}}}
\end{minipage}
\caption{Contour plot of $1/(1+ |t_0|)$ (up) 
and $1/(1+|z_1|)$ (down)
for $k=0.25$ (left) and $k=0.5$ (right).
The vertical axis stands for 
$-\frac{\pi}2 \le \rm{Im}(\theta)\le \frac{\pi}4$
and the horizontal one for 
$0 \le e^{\rm {Re} (\theta)} \le 6$.
The moving zeros of $t_0$ and $z_1$ are pointed 
with arrows.}
\label{2-singular}
\end{figure}

It is to noted that all the zeros of $t_0$ 
lie at Im($\theta$)=$\pm\pi/2$
instead of Im($\theta$)=$\pm\pi/4$.
In addition, there are no zeros of $z_1$ 
in this domain.
(Note that pseudo energy $\epsilon_A$ 
is symmetric under $\theta \to \theta^*$ 
when $\chi$ is real).

Next, the singularity positions are traced
with the complexified $\chi$.
$\chi$ is put as $\chi \to \frac{\pi}3 
- (\frac\pi{30}) e^{i \pi k}$
and $k$ is varied $0 \to 1$.
The numerical result is given in Fig.~\ref{2-singular}.
As $k$ increases, one of the $t_0$ zeros 
(with the smallest 
value of real rapidity) goes up.
When $k= 0.25$, 
one zero of $z_1$ with the same real rapidity 
appears at Im($\theta)=-\pi/2$ 
and moves up as $k$ increases.
As $k$ reaches 0.5, the zero of $z_1$ 
comes to Im($\theta)=-\pi/4$.

In fact the positions of zeros 
of each species are related with 
the positions of zeros of other species 
according to the Y-system Eq.~(\ref{Y-system}).
As the singularity of one species 
moves around, the corresponding singularities  
of other species follow the trace. 

\begin{figure}[h]
\begin{minipage}[t]{7.5cm}
{\scalebox{0.4} {\includegraphics{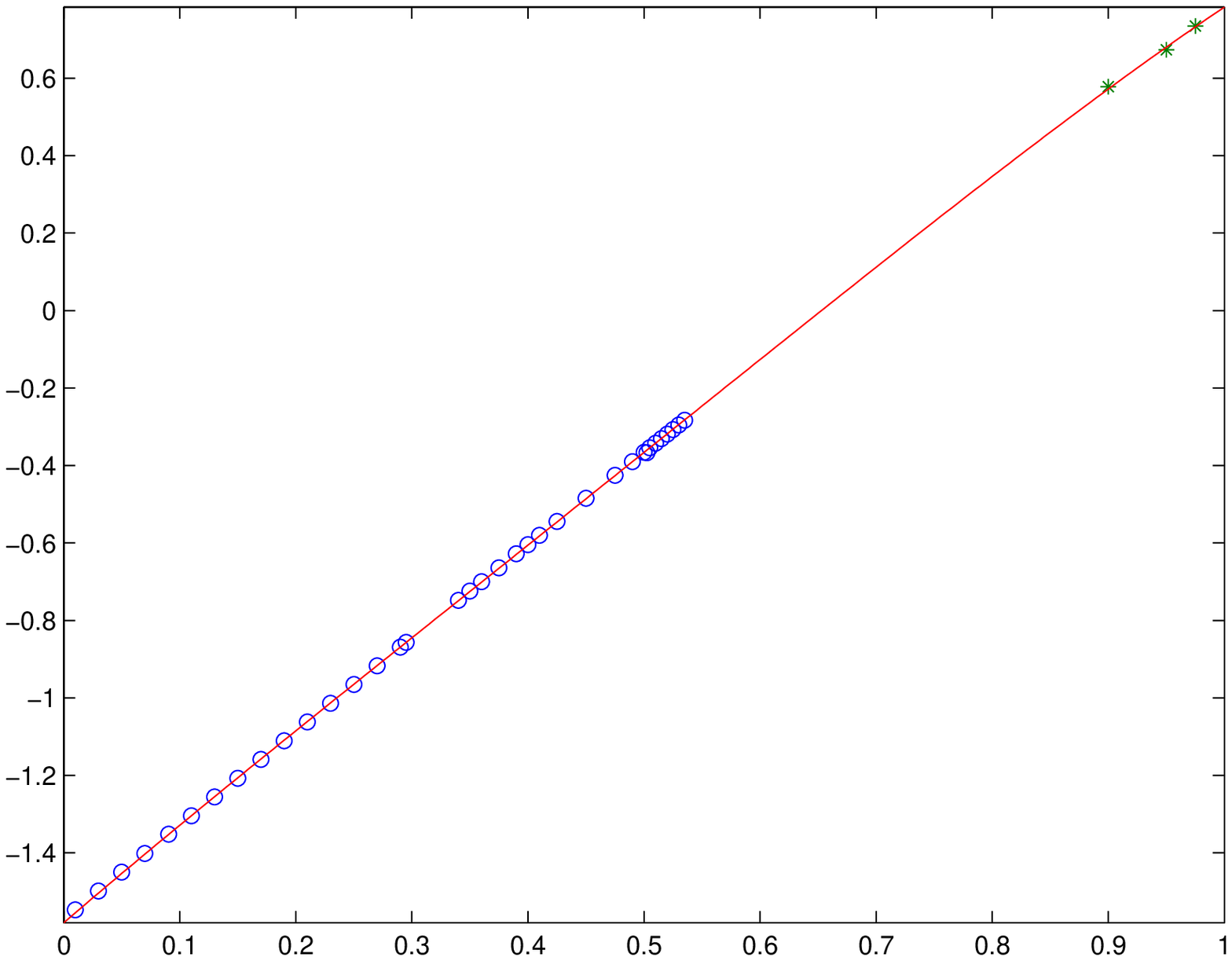}}}
\caption{$t_0$ crossing singularity position:
$e^{\rm{Re}(\theta)}$ v.s. $k$.
Data (denoted as o) for less than $k=0.6$
are taken from the original TBA.
Data (denoted as $*$) for close to $k=1$
are taken from the modified TBA.}
\label{2-imag-crossing}
\end{minipage}
\ $\qquad$ \
\begin{minipage}[t]{7.5cm}
{\scalebox{0.4} {\includegraphics{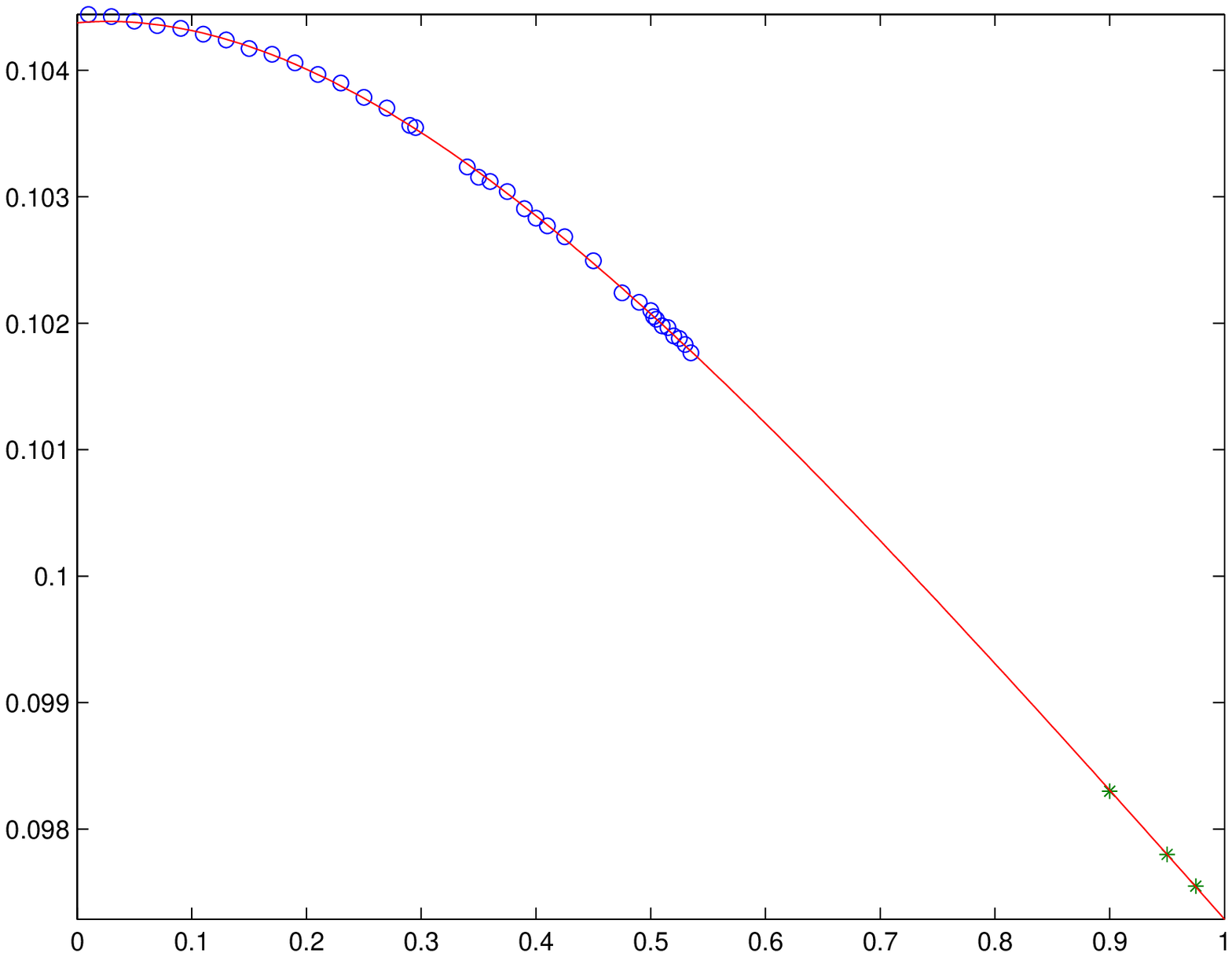}}}
\caption{$t_0$ crossing singularity position:
Im ($\theta$) v.s. $k$.
Data (denoted as o) for less than $k=0.6$
are taken from the original TBA.
Data (denoted as $*$) for close to $k=1$
are taken from the modified TBA.}
\label{2-real-crossing}
\end{minipage}
\end{figure}

The position of $t_0$ zero 
is drawn in Figs.~(\ref{2-imag-crossing}, \ref{2-real-crossing}).
The data is taken up to $k=0.5$ without any difficulty.
However, beyond $k=0.5$ numerical work becomes harder 
due to the numerical instability. We use the 
method using the ``damped'' iterative 
method used in \cite{DT}
to easy this instability up to near $k=0.56$.
Beyond that point numerical instability 
does not allow any further computation.
Neverthelss, the data clearly 
indicates that 
the imaginary position changes linearly in $k$. 

Now, it should be noted that 
as the $t_0$ zero crosses the real rapidity line, 
we cannot use the TBA Eq.~(\ref{m0sG-TBA}) any longer.
This is because the moving 
singularity pushes up the real integration
line of the convolution
and forces to modify the TBA.
Denoting the position of the $t_0$ zero as 
$\tilde \theta$, we have 
\beaq
\epsilon_{A} (\theta)  &=& 2 m_{A}\, r\, e^\theta 
+ \ln S_{A\,0}(\theta -\tilde \theta ) 
-\frac1{2\pi} \sum_{B=0,1} 
-\!\!\!\!\!\! \int_{-\infty}^\infty d \theta' 
\phi_{A\,B}(\theta -\theta') \tilde L_{B}(\theta')\,.
\label{bTBA-1}
\eeaq
Here the branch-cut contribution is used as follows.
\bea
-\frac1{2\pi} \int d\theta' \phi_{A0} (\theta -\theta') 
\tilde L(\theta') 
\quad\to\quad    \ln S_{A0} (\theta - \tilde \theta) 
-\frac1{2\pi} -\!\!\!\!\!\!\int\,
d\theta' \phi_{A0} (\theta -\theta') 
\tilde L(\theta') \,.
\eea
 
\begin{figure}[h]
\begin{minipage}{7.5cm}
{\scalebox{0.45} {\includegraphics{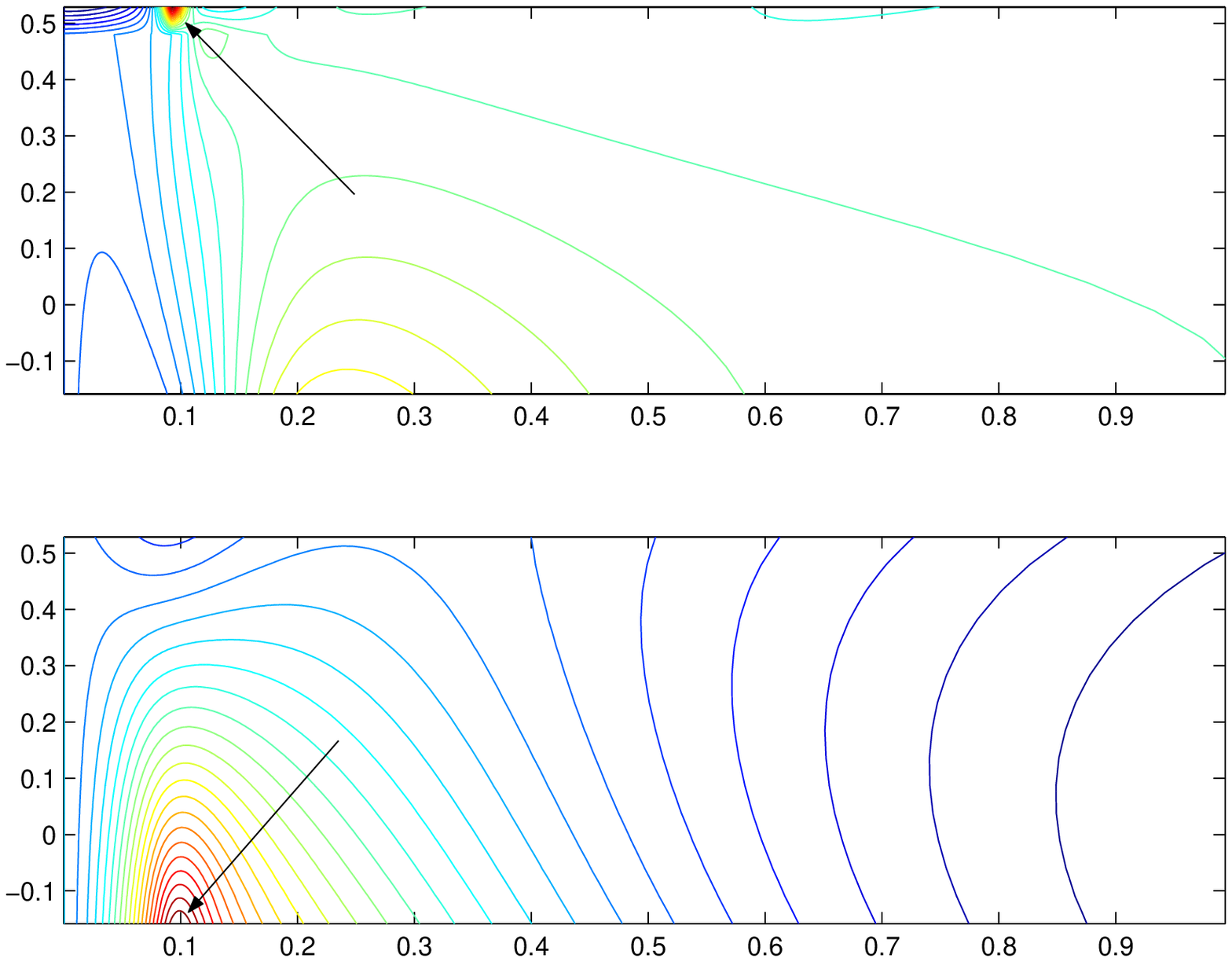}}}
\caption{Position of  $t_0$ zero (up) and $z_1$ (down)
for $k=0.9$.
Vertical axis stands for 
$ 0.58 -\frac{\pi}4 \le \rm{Im}(\theta)\le 0.58$
and horizontal one for 
$0 \le e^{\rm {Re} (\theta)} \le 1$.
The moving zeros of $t_0$ and $z_1$ are pointed 
with arrows.}
\label{2-aftercrossing-1}
\end{minipage}
\ $\qquad$ \
\begin{minipage}{7.5cm}
{\scalebox{0.45} {\includegraphics{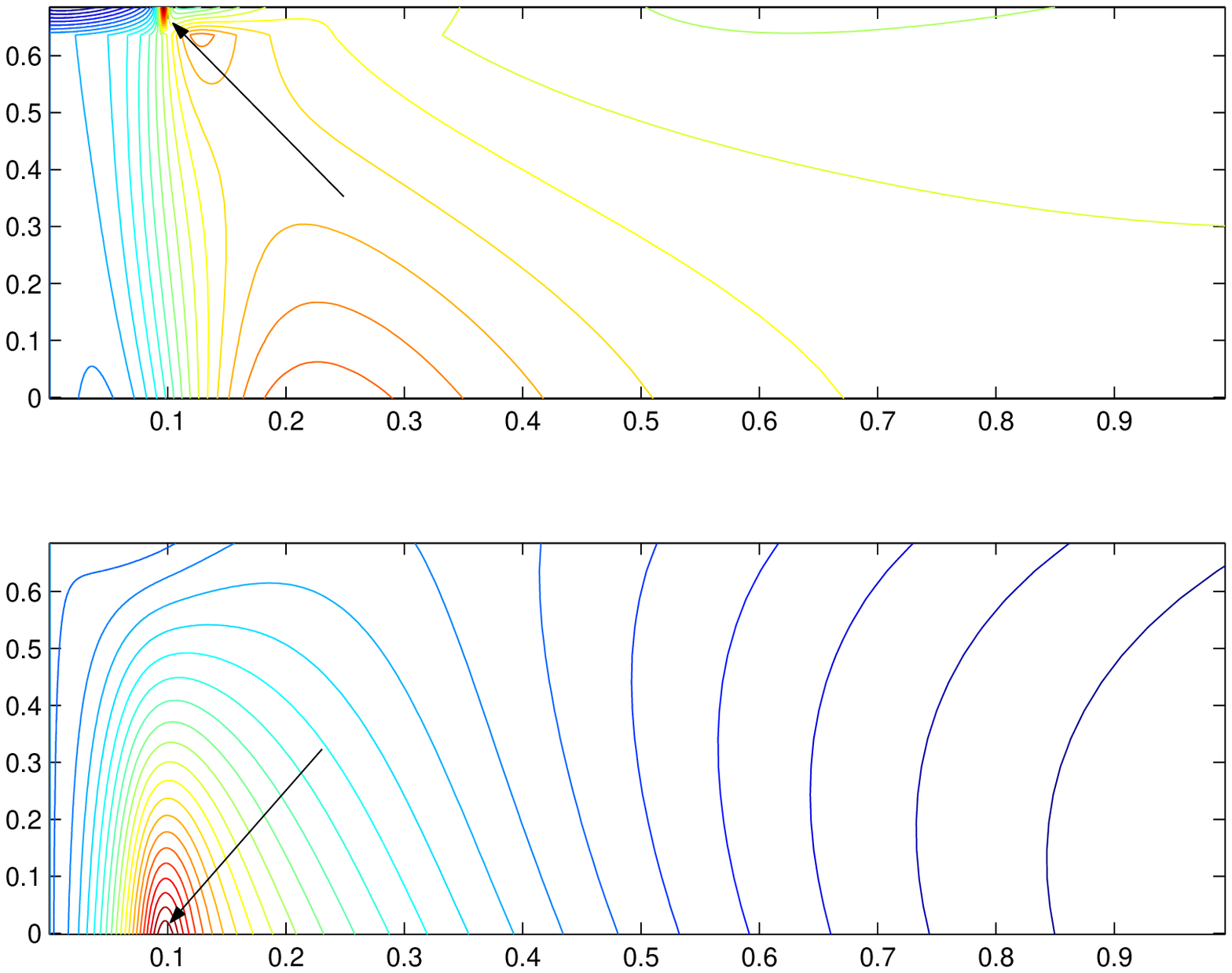}}}
\caption{Position of  $t_0$ zero (up) and $z_1$ (down)
for $k=0.975$.
Vertical axis stands for 
$ 0.73-\frac{\pi}4 \le \rm{Im}(\theta)\le 0.73$
and horizontal one for 
$0 \le e^{\rm {Re} (\theta)} \le 1$.
The moving zeros of $t_0$ and $z_1$ are pointed 
with arrows.}
\label{2-aftercrossing-2}
\end{minipage}
\end{figure}

Using this modified TBA Eq.~(\ref{bTBA-1}) 
we obtain Figs.~(\ref{2-aftercrossing-1}, \ref{2-aftercrossing-2})
for the singularity structure.
The figures show that $t_0$ zero and 
$z_1$ zero have the same real rapidity 
($e^{{\rm Re}\theta} \approx 0.1$) but with
the imaginary rapidity $\pi/4$ apart.
In addition, the position of $t_0$ zero 
in Figs.~(\ref{2-imag-crossing}, \ref{2-imag-crossing})
(denoted as $*$) 
ends up at the imaginary rapidity $\pi/4$.

As $k$ approaches 1, $\chi$ becomes real and 
the $t_0$ zero reaches Im($\theta)=\pi/4$.
At the same time the zero of $z_1$ reaches 
the real rapidity axis, Im($\theta)=0$. 
Therefore, at real $\chi = \pi/3 + \pi/30 > \pi/3$,
the $z_1$ zero
in the convolution integration
modifies TBA again. 
For real $\theta$  the pseudo energy becomes
\beaq
\epsilon_{A} (\theta)  &=& 2 m_{A}\, r\, e^\theta  
+ \ln S_{A\,0}(\theta -\tilde \theta ) 
-\frac 12  \ln S_{A\,1}(\theta -\tilde \theta ) 
-\frac1{2\pi} \sum_{B=0,1}
-\!\!\!\!\!\!\int_{-\infty}^\infty d \theta' 
\phi_{A\,B}(\theta -\theta') \widetilde L_{B}(\theta')
\nn\\ 
&=& 2 m_{A}\, r\, e^\theta 
+ \frac12 \ln \frac{S_{A\,0}(\theta -\tilde \theta )} 
{S_{A\,0}(\theta -\tilde \theta^* )} 
-\frac1{2\pi} \sum_{B=0,1} 
-\!\!\!\!\!\!\int_{-\infty}^\infty d \theta' 
\phi_{A\,B}(\theta -\theta') \widetilde L_{B}(\theta')\,.
\label{bTBA-2}
\eeaq 
In the last identity
the bootstrap relation of the scattering matrix
is used:
\bea
S_{A0}(\theta -i\pi/4) 
S_{A0}(\theta +i\pi/4) 
=S_{A1}(\theta)\,, \qquad  A=1,2 \,.
\eea
Explicitly,
\bea
S_{00}(\theta)=i \tan \bigg( \frac{\theta + i\pi/2}2\bigg)\,,
\quad
S_{11}(\theta)=-\bigg( S_{00}(\theta) \bigg)^2\,,
\quad
S_{10}(\theta)=\frac{1- \sqrt2 \cosh(\theta +i\pi/2)}
{1+\sqrt2 \cosh(\theta +i\pi/2)}\,.
\eea

Note that the convolution integration 
in Eq.~(\ref{bTBA-2}) refers to the
principal value due to the singularity 
of $\widetilde L_{1}(\theta')$ at $\theta=\theta_p$. 
The numerical integration of the convolution 
can be done by shifting the contour 
integration. 
Another way of avoiding this numerical instability is 
to use TBA with the reduced kernel.
Using the kernel relation (\ref{kernelrelation}), we have
\beq
\epsilon_{A'}(\theta)
=
D_{A'}(\theta) + P_{A'}(\theta)
+ \sum_{B'=0,1,+} \mathcal I_{A'\,B'} \,
-\!\!\!\!\!\int_{-\infty}^{\infty} d\theta'\,
K(\theta-\theta') \bigg(
\widetilde{L_{B'}} 
+\epsilon_{B'} -D_{B'} - P_{B'}\bigg)(\theta') 
\label{bTBA-r}
\eeq
where $ P_{A'} (\theta)
=
\frac12 \ln \frac{S_{A'\,0}(\theta -\tilde \theta )} 
{S_{A'\,0}(\theta -\tilde \theta^* )} $
and $D_{A'} = 2 m_{A'} r e^{\theta}$. 
Note that $(\widetilde{L_1} 
+\epsilon_1)(\theta')$ vanish at $\theta'=\theta_p$.

\begin{figure}[ht]
\center{
{\scalebox{0.5} {\includegraphics{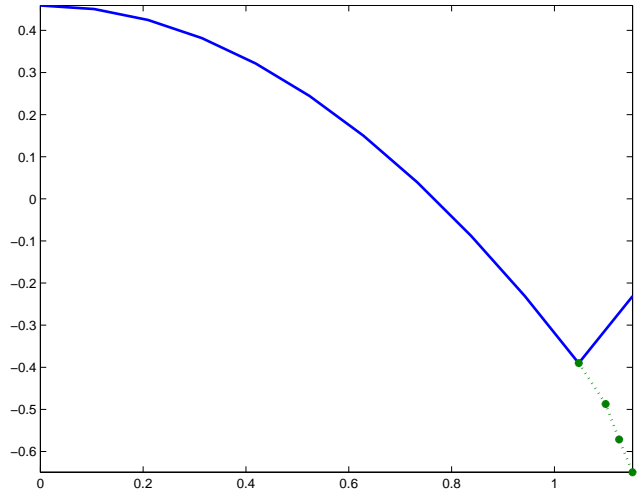}}}}
\caption{$c_{\rm eff}$ vs. $\chi $ when  $\lambda=2$ 
and $\vartheta= 0 $:
Solid line is obtained from the original TBA
and dotted one from the modified TBA.}
\label{2-ceff} 
\end{figure}

Now, the crossing singularities 
contribute to the free energy:
\beq - \frac \pi{24 r} c_{\rm eff}
= -\frac{m_0}{\sqrt2} \, e^{\theta_p } 
- \frac1{2\pi}
 -\!\!\!\!\!\!\int_{-\infty}^\infty d \theta
\,\sum_{A=0,1} m_{A } \, e^\theta \, \tilde L_A(\theta)\,,
\label{modfiedceff}
\eeq
where $\theta_p$ is the real part of $\tilde \theta$.
This singularity position should satisfy
the constraint equation,
\beq
t_0(\theta_p \pm i \frac{\pi}4) = z_1(\theta_p) =0
\eeq
Thus obtained $c_{\rm eff}$ is compared with 
the original result in Fig.~\ref{2-ceff}.  
The smooth behavior of $c_{\rm eff}$ 
in Fig.~\ref{2-ceff} indicates that  
the modified TBA is in the right track.
 
Finally, we remark that 
the modified TBA~(\ref{bTBA-2}, \ref{bTBA-r}) 
is also written in another form 
if the kernel relation (\ref{kernelrelation}) is used
for the S-matrix:
\beq
\epsilon_{A'}(\theta)= D_{A'}(\theta)
+ \sum_{B'=0,1,+} \mathcal I_{A'\,B'} \,
\Bigg( \Xi (\theta, \tilde \theta)
+  \int_{-\infty}^{\infty} d\theta'\,
K(\theta-\theta') (\widetilde{L_{B'}} 
+\epsilon_{B'} -D_{B'})(\theta') \Bigg)
\label{bTBA-rr}
\eeq
where
\bea
i\,\Xi (\theta, \tilde \theta)
= \tan^{-1}(\tanh(3(\theta-\tilde \theta)/2))
- \tan^{-1}(\tanh(3(\theta-\tilde \theta + i \pi/2 )/2))\,.
\eea
This TBA satisfies the same Y-system in Eq.~(\ref{Y-system}) 
with the fugacities replaced with the massless one
(h=4 when $\lambda=2$). 
Redefining $Y$'s, 
$y_0(\theta) =Y_0(\theta) /\sqrt{\lambda^d (\theta)}\,$,
$\,\,y_1(\theta) =Y_1(\theta) /\lambda^1(\theta)\,$ 
and $\omega = i \pi/4$, we may put into this form.
\beaq
y_0(\theta + \omega)\,y_0(\theta - \omega) 
&=& 1 + y_1(\theta)
\nonumber\\
y_1(\theta + \omega)\,y_1(\theta - \omega) 
&=& 1 + \frac{\lambda^0}{\sqrt{\lambda^d}}
y_0(\theta) + \bigg(y_0(\theta)\bigg)^2 
\eeaq
where fugacity relations for $\lambda=2$ are also used,
\bea
\lambda_{\alpha\beta}^d (\theta + \omega)\,
\lambda_{\alpha\beta}^d (\theta - \omega)
&=& \bigg( \lambda_{\alpha\beta}^1 (\theta) \bigg)^2\\
\lambda_{\alpha\beta}^1 (\theta + \omega)\,
\lambda_{\alpha\beta}^1 (\theta - \omega)
&=& \lambda_{\alpha\beta}^d (\theta) \,.
\eea
Note that since $ \lambda^0(\theta)/\sqrt{\lambda^d(\theta)}$
is $ \frac{(h+2)\pi}{h} =  \frac{3\pi}2$-periodic in $\theta$, 
so are $y_0 (\theta)$ and $y_1 (\theta)$ as in the bulk case.

\section{Conclusion} 

We presented the (R-channel) TBA for the (massless) boundary
sine-Gordon theory with coupling parameter 
$\lambda =$ positive integer, which incorporates
the violation of the topological charge.  
With this TBA, the boundary parameter effect 
on the effective central charge 
is investigated. 

This TBA, however, doest not reflect
the boundary parameter condition faithfully
beyond the cusp point  $\chi =b^2 \pi$. 
We have demonstrated that this unpleasant
feature is due to the branch singularity 
crossing in the TBA equation. 
Numerical analysis show that 
near the cusp point of $c_{\rm eff}$ 
some of 
the singularities of TBA crosses the real 
rapidity. 
It turns out that  
only one singularity for each species 
changes its position in the complex plane.
Especially, its imaginary position moves 
``linearly'' with 
the phase of the complexified  
boundary parameter $\chi$ near the cusp. 
We expect this singularity crossing feature will
be very universal.
When $\lambda=2$ ($h=4$ 
and periodicity $\frac32$ in unit of $\pi$), 
the imaginary parts of the 
original singularity positions are
$ \pm \frac24, \frac34$.
After singularity crossing some of the positions
change into 
$0, \pm \frac14$.
This leads into consideration for other $\lambda$'s
($h=2\lambda$).
From the original singularity positions 
$ \pm \frac2{h}, \pm \frac3{h}, \cdots,
\pm \frac{\lambda}{h}, \frac{\lambda+1}{h}$ 
with periodicity $\frac {h+2}h$,
some of the singularities will presumably 
rearrange into 
$0, \pm \frac1{h}, \pm \frac2{h}, \cdots,
\pm \frac{\lambda-1}{h}$.
This expectation 
needs to be checked further
as well as the behavior for various parameter range.
The scale dependence of the massive TBA 
is also not well understood.  
This issues will be carried on 
in a separate paper. 

This analysis is also expected to hold 
for an arbitrary coupling constant case.
However, the TBA of massless/massive
boundary sine-Gordon theory with arbitrary coupling
is not feasible at this moment
since the bulk Hamiltonian eigenstates 
will result in the infinitely 
coupled TBA equations \cite{sGTBA}. 
Instead of this approach,
DDV type equation is expected to be more suitable,
which does not impose the string hypothesis 
for the structure of the roots of Bethe ansatz 
\cite{ddv}. 

It is known \cite{alyosh-phase} that 
the sign change effect of the boundary term 
of boundary Liouville theory 
and boundary sinh-Gordon model 
\cite{bshG} can be explained 
using the analytically continued boundary parameter
only to a certain range.  
Nevertheless, the boundary Lagrangian 
phase difference is also known \cite{alyosh-phase} 
to induce the branch singularity crossing 
similarly as in this boundary sine-Gordon model.
Furthermore, the
integrable boundary ADE-affine Toda theories \cite{afftoda} 
have discrete boundary conditions,
(+), ($-$) and Neumann condition.
Among the three, ($-$) boundary condition
is not yet fully understood.
It remains to be seen 
if there is any relation between 
the relative sign change of the boundary term
$\mu_B \to -\mu_B$ and the exicted state contribution 
to $c_{\rm eff}$.\\

{\sl Note added}: While our manuscript being revised, 
a preprint by Caux  et.\ al.\ (cond-mat/0306328) appeared
where similar singularity structure information 
at the cusp point 
is used for the analysis of the finite size effect. 

\section*{\bf Acknowledgement}

We thank C. Ahn, P. Dorey, K. Moon and R. Tateo 
for invaluable discussions and KIAS for hospitality.
CR thanks Al. Zamolodchikov for 
explaining his work on
the singularity crossing in boundary sinh-Gordon model,
H. Saleur for informing their singularity conjecture 
of TBA 
and Durham University 
where the revision of the manuscript was made. 
This work is supported in part 
by the Basic Research Program of the Korea Science 
and Engineering Foundation Grant number R01-1999-00018-0(2002) and by Korea Research Foundation 2002-070-C00025.

\end{document}